# Bits through ARQs

Krishnan Eswaran, Michael Gastpar, Kannan Ramchandran

Dept. of Electrical Engineering and Computer Sciences

University of California, Berkeley

Berkeley, CA 94720

E-mail: {keswaran, gastpar, kannanr}@eecs.berkeley.edu

### Abstract

A fundamental problem in dynamic frequency reuse is that the cognitive radio is ignorant of the amount of interference it inflicts on the primary license holder. A model for such a situation is proposed and analyzed. The primary sends packets across an erasure channel and employs simple ACK/NAK feedback (ARQs) to retransmit erased packets. Furthermore, its erasure probabilities are influenced by the cognitive radio's activity. While the cognitive radio does not know these interference characteristics, it can eavesdrop on the primary's ARQs. The model leads to strategies in which the cognitive radio adaptively adjusts its input based on the primary's ARQs thereby guaranteeing the primary exceeds a target packet rate. A relatively simple strategy whereby the cognitive radio transmits only when the primary's empirical packet rate exceeds a threshold is shown to have interesting universal properties in the sense that for unknown time-varying interference characteristics, the primary is guaranteed to meet its target rate. Furthermore, a more intricate version of this strategy is shown to be capacity-achieving for the cognitive radio when the interference characteristics are time-invariant.

## I. INTRODUCTION

Systems often need to be designed so that they do not disrupt pre-existing systems with which they interact. This backwards compatibility problem is a central issue in the study of cognitive radio systems. A cognitive radio is a device that can sense and adjust its power, frequency band, etc. to peacefully coexist with other radios with which it shares spectrum [1]. The FCC and international regulatory bodies are considering modifying their rules to allow for such systems to occupy unlicensed bands or to share bands with licensed, predesigned communication systems. These licensed users are often called primaries, legacy systems, or incumbents.

The aim of this paper is to study sharing spectrum with legacy systems, in which the backwards compatibility problem arises. One potential solution is to transmit on a band that is currently unoccupied and to leave that band once a primary is detected. For these "detect-and-avoid" systems, one research aim is to understand the feasibility of detecting the presence of a primary system subject to noise uncertainty and quantization effects [2], [3], [4].

A different approach is for the cognitive radio to occupy bands on which the primary is already active but in such a way as to mitigate the interference generated on the primary system. Two such information-theoretic models have been introduced to study cognitive radio and spectrum sharing systems. The first is sometimes called the cognitive radio channel [5], [6], [7], [8], [9], [10], [11]. This channel is a variation on the two-user interference channel [12], [13], [14], [15] with the modification that the cognitive radio (one of the transmitters) knows the message that the primary (the other transmitter) will send. Among these papers, Devroye, Mitran, and Tarokh [5] as well as Jovičić and Viswanath [6] consider a Gaussian scenario in which the primary's strategy can be thought of as a fixed, predesigned legacy system. Specifically, they show that for their setup, there is an optimal achievable strategy that enables the primary to continue using a point to point Gaussian codebook. The result highlights the fact that in cognitive radio problems, one may not have the flexibility to modify the primary's design and must instead design the cognitive radio in such a way that the primary continues to meet its target performance. The second approach is to consider the capacity of systems with a constraint on the interference power generated at certain locations. The assumption is that the primary systems that occupy these locations will be able to handle this level of interference [16], [17].

We take inspiration from these two models in the following example, which forms the starting point of the current investigation. It is a model of the practically most interesting case, where the cognitive transmitter is close to the primary receiver, thus creating substantial interference. For simplicity, we assume the cognitive radio's receiver is shadowed from the primary transmitter, thereby avoiding interference from that system. An illustration of the setup is given in Figure 1.

*Example 1:* Suppose the primary sends packets across an erasure channel and receives feedback from its receiver to retransmit the packet or send the next one. The cognitive radio, on the other hand, has a noiseless channel to its receiver with $P + 1$ channel inputs divided into two classes: a *silent symbol* $x_{\text{off}}$ results in a successful receipt of the primary's transmitted packet, and the remaining $P$ *transmit symbols* cause the primary's packet to be erased. Suppose the primary wants a guaranteed rate of $\frac{1}{2}$; that is, one packet should be successfully received per two transmissions on average. By simply alternating channel

 

uses between the silent symbol and sending information with the $P$ transmit symbols, the cognitive radio guarantees the primary rate $1/2$ target and can itself achieve a rate of

$$\frac{1}{2} \log P \ , \tag{1}$$

where $P + 1$ is the number of channel input symbols available to the cognitive radio.

In the spirit of the previous work, Example 1 considers a primary that is unaware of the cognitive radio. However, Example 1 makes the more dubious assumption that the cognitive radio knows what the primary's erasure probabilities are for its two classes of inputs. As a result, the strategy presented is not robust to deviations from the erasure probabilities provided in the example. For instance, if the primary's erasure probability for a silent symbol $x_{\text{off}}$ is $\epsilon_0 > 0$, the strategy outlined will not allow the primary to meet its rate $1/2$ target. The issue is that the cognitive radio will not be able to directly estimate the interference it creates for the primary. Such estimates are generally obtained by training via pilot symbols, but the primary receiver is unlikely to train with the cognitive radio transmitter. However, certain kinds of 1-bit feedback have been shown to be sufficient for beamforming [18], [19]. We adopt this insight as we build on Example 1 by introducing both an uncertainty and sensing component to the problem.

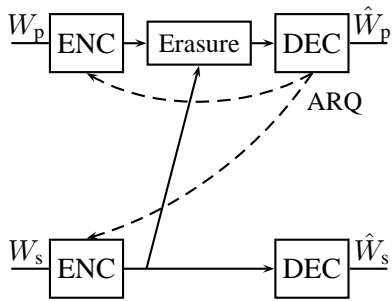

Fig. 1. An example of the type of channel model for our cognitive radio system in which the primary has message $W_{\text{p}}$ and the cognitive radio has message $W_{\text{s}}$. The cognitive radio transmitter can listen to the ARQ feedback the primary receiver sends to its transmitter to adapt its transmission rate and reduce interference on the primary system.

*Example 1, continued:* The cognitive radio's silent symbol now induces an erasure probability $\epsilon_0 < 1/2$, and its transmit symbols induce an erasure probability of $\epsilon_1 > 1/2$, both of which are unknown to the cognitive radio. However, the cognitive radio transmitter can sense the primary's ARQs, which we will denote with the indicator random variables $A_k$ when the primary's $k$-th transmission is received. Figure 1 shows a schematic block diagram of this setup.

The cognitive radio's strategy is as follows. If the primary is exceeding its rate target at time $k$, the cognitive radio sends one of its transmit symbols based on its message. This happens when $\frac{1}{k} \sum_{i=1}^{k} A_k \geq \frac{1}{2}$. Otherwise, the cognitive radio sends its silent symbol. Let $\tau_k$ be the indicator function that the cognitive radio sends a transmit symbol at time $k$. Thus, $\mathbb{P}(A_k = 1 | \tau_k) = 1 - \epsilon_{\tau_k}$. Then the cognitive radio's rate at time $n$ is

$$\frac{1}{n} \sum_{k=1}^{n} \tau_k \log P \ . \tag{2}$$

Note that for $\epsilon_0 = 0$, $\epsilon_1 = 1$, this strategy is as good as the one outlined in Example 1.

What can we say about the rate for the primary and cognitive radio in Example 1? Let $S_0 = 0$ and $S_k = S_{k-1} + (A_k - 1/2)$ represent the difference between the number of packets the primary has received by time $k$ and its targeted number of packets by time $k$ based on a target packet rate of $\frac{1}{2}$. Suppose the $A_k$ are independent in $k$. Then $S_k$ is a positive recurrent Markov chain and is nonnegative if and only if $\tau_k = 1$, which can be verified by confirming that its stationary distribution is

$$\pi_{i/2} = \begin{cases} \frac{(2\epsilon_1 - 1)(1 - 2\epsilon_0)}{2\epsilon_1(\epsilon_1 - \epsilon_0)} \left( \frac{1 - \epsilon_1}{\epsilon_1} \right)^i & i \geq 0 \\ \frac{(2\epsilon_1 - 1)(1 - 2\epsilon_0)}{2(1 - \epsilon_0)(\epsilon_1 - \epsilon_0)} \left( \frac{\epsilon_0}{1 - \epsilon_0} \right)^{-i+1} & i \leq 0 \end{cases} \ . \tag{3}$$

We can make the following statement.

*Fact:* Suppose $S_0$ is distributed according to $\pi$. Then for all $k \geq 1$,

$$\mathbb{P}(\tau_k = 1) = \sum_{i=0}^{\infty} \pi_{i/2} = \frac{1/2 - \epsilon_0}{\epsilon_1 - \epsilon_0} \ . \tag{4}$$

The fact allows us to get a handle on the cognitive radio's rate. Furthermore, the primary's expected rate is

$$k^{-1} \sum_{i=1}^{k} \sum_{j=0,1} \mathbb{P}(A_i = 1 | \tau_i = j) \mathbb{P}(\tau_i = j) = \frac{1}{2} \ . \tag{5}$$

Note that this strategy does not depend on the cognitive radio knowing the values $\epsilon_0$ and $\epsilon_1$ a priori. However, the cognitive radio does know the primary's rate target, which is $1/2$ in this example. In the remainder of the paper, we assume the primary's rate target is known in advance to the cognitive radio, but the primary's erasure probabilities are unknown.

In this work, we consider optimal coding strategies for the case in which the primary is a packet erasure system as described in Example 1.[1] For the channel of the cognitive radio, we consider a more general class of (noisy) channels. As we show, the primary can meet its target rate even if the cognitive radio is active for a certain fraction of channel uses. This *interference budget* available to the cognitive radio, while unknown a priori, can be estimated via the primary ARQs and rate target, which are known at the cognitive radio encoder. One can determine the capacity of the cognitive radio in terms of this interference budget, which we call the rate-interference budget (RIB) tradeoff function. We show an achievable strategy for the general case in which the primary's packet erasure probabilities can fluctuate and find a matching converse for the RIB function when they do not.

---

[1] This formulation lends itself well to many spectrum sharing problems in which the primary is a separately designed system and whose exact implementation is partially obscured from the cognitive radio.

In Section II, we define the problem we are considering precisely, including the channel model for the cognitive radio and the allowable coding strategies that the cognitive radio can adopt. These strategies force the cognitive radio to provide guarantees about the primary's rate that do not depend on the time horizon that the cognitive radio uses to measure its own rate (*horzion-independence* condition) and force it to be robust to fluctuations in the primary's packet erasure probabilities (*robustness* condition).

In Section III, we show how to refine the strategy from Example 1 to provide such guarantees that also allow positive rate for the cognitive radio, which leads to two new strategies: the *fixed-codebook protocol* and the *codebook-adaptive protocol*. In Section IV, we present a converse when the erasure probabilities are time-invariant, which matches the rates achievable by the codebook-adaptive protocol proposed in Section III. Section V revisits Example 1 in the introduction and considers new ones. Section VI concludes the paper with a discussion of our contributions and future work.

## II. PROBLEM SETUP AND MAIN RESULT

Capital letters $X, Y, Z$ represent random variables and calligraphic letters $\mathcal{X}, \mathcal{Y}, \mathcal{Z}$ denote finite sets. We will focus on discrete memoryless channels in this work, but potential extensions to Gaussian channels will be discussed in Section VI. For convenience, $p(x)$ is the probability distribution of $X$ at $x$. Similarly, $p(y|x)$ is the conditional probability distribution of $Y$ at $y$ given $X = x$. Notation for entropy $H(X)$, mutual information $I(X; Y)$, etc. are consistent with the notation of Cover and Thomas [20].

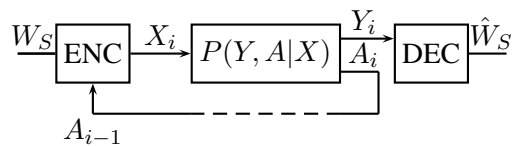

Fig. 2. Equivalent channel model from the cognitive radio's perspective. $A_i$ is an indicator random variable: $A_i = 1$ means that the packet sent by the primary at time $i$ was successfully received.

## A. Equivalent Channel Model

As a legacy ARQ system, the primary is assumed to have the following fixed strategy. At time $i$, it sends a packet to its receiver and receives feedback $A_i$ to indicate whether the packet was erased or successfully received ($A_i = 0$ or $1$, respectively). If the packet is erased, the primary retransmits the same packet at time $i + 1$. If the packet is successfully received, the primary transmits a new packet at time $i + 1$. Thus, we will refer to $A_i$ as the *primary's ARQ feedback*.

Since the primary's strategy is fixed, we now have to design the cognitive radio's strategy. Figure 2 illustrates this problem; the primary merely appears as a constraint on the cognitive radio in the shape of $A_i$. That is, in addition to communicating, the cognitive radio must also control its channel inputs to guarantee the primary's rate, i.e. such that to the first-order[2],

$$k^{-1} \sum_{i=1}^{k} A_i \geq R_{\mathrm{p}} \ , \tag{6}$$

where $R_{\mathrm{p}}$ is the desired and prespecified performance of the primary system. Furthermore, this control must be robust to fluctuations in the channel between the cognitive radio transmitter and primary receiver. Thus, the primary's ARQ feedback provides a means for the cognitive radio to apply this control.

## B. Channel Model and Coding

We now consider the DMC with feedback from Figure 2 in more detail. Let $\mathcal{X} = \{x_{\mathrm{off}}, 1, \ldots, |\mathcal{X}| - 1\}$ be the channel inputs. Then at time $i$, the conditional distribution of the channel output $Y_i$ and primary's ARQ $A_i$ given $X_i = x$ can be expressed as

$$p(y_i, a_i | x_i = x) = p(y_i | x_i = x) \cdot \epsilon_{x,i} \cdot \exp\left(a_i \cdot \log \frac{1 - \epsilon_{x,i}}{\epsilon_{x,i}}\right) \ . \tag{7}$$

We assume that the sequences $\{\epsilon_{x,i}\}_{i=1}^{\infty}$, for $x \in \mathcal{X}$ are unknown at the encoder and decoder. For simplicity, we assume

$$\epsilon_{x_{\mathrm{off}},i} = \epsilon_0 \tag{8}$$

does not depend on $i$, which assumes that the primary's channel is fixed when the cognitive radio is silent.

However, allowing $\epsilon_{x,i}$ to vary with $i$ for $x \neq x_{\mathrm{off}}$ reflects uncertainty about the amount of interference the cognitive radio is generating on the primary. We will assume that for all $x \neq x_{\mathrm{off}}$ and $i = 1, 2, \ldots$,

$$\epsilon_{x,i} > \epsilon_0 \tag{9}$$

---

[2]Second order issues and tight delay constraints are discussed in Section VI.

                                                                      

In the remainder of this paper, we make the technical assumption that there is a known constant $\nu > 0$ such that for all $i$, $R_{\mathrm{p}} < 1 - \epsilon_0 - \nu$. This assumption enables the primary to tolerate some interference from the cognitive radio while guaranteeing the cognitive radio achieves a positive rate.

The definition of the rate and capacity for the secondary are complicated by the fact that the number of channel uses depends on the realizations of $\epsilon_{x,i}$. Therefore, we need to be precise on what is meant by messages. We define the set of possible messages to be the set of binary sequences $\{0,1\}^{nC_{\max}}$, where $C_{\max} = \log\min\{|\mathcal{X}|, |\mathcal{Y}|\}$. Let $W_k$ be the first $k$ bits of the message and $W = W_{nC_{\max}}$.

An $(n, f^n, g)$ *code* (we call $n$ the blocklength) consists of a set of encoding functions $f_i : \{0,1\}^{i-1} \times \{0,1\}^{nC_{\max}} \to \mathcal{X}$ for $i = 1, 2, \ldots, n$,

$$X_i = f_i(A^{i-1}, W) \ , \tag{10}$$

and decoding function $g : \mathcal{Y}^n \to \{0,1\}^{nC_{\max}}$

$$\hat{W} = g(Y^n) \ . \tag{11}$$

A *strategy* is a sequence of $(n, f^n, g)$ codes indexed by $n$ on the positive integers $n = 1, 2, \ldots$.

Strategies must respect the primary's rate target, so the following definition restricts the type of strategies we allow.

A strategy is *valid* if for all $\nu > 0$ and for $k \leq n$ in each $(n, f^n, g)$ code,

$$\mathbb{P}\left(k^{-1} \sum_{i=1}^{k} A_i \leq R_{\mathrm{p}}\right) \leq K_{1, R_{\mathrm{p}}, \nu, k} e^{-k \cdot K_{2, R_{\mathrm{p}}, \nu}} \ , \tag{12}$$

where the constants $K_{1, R_{\mathrm{p}}, \nu, k} < \infty$, $0 < K_{2, R_{\mathrm{p}}, \nu} < \infty$ depend only on the fat in the system $\nu$ and target rate $R_{\mathrm{p}}$, and the right side of (12) goes to 0 as $k \to \infty$.

Note that a valid strategy imposes two restrictions. First, the convergence of the primary's rate should not depend on the blocklength of a strategy (*horizon-independence* condition). Second, the convergence of the primary's rate should be the same irrespective of $\epsilon_0, \{\epsilon_{x,i}\}_{i=1}^{\infty}, x \in \mathcal{X} - \{x_{\mathrm{off}}\}$ (*robustness* condition). For a given valid strategy, we will use the notation $\hat{W}^n$ to denote the decoded output for its code of blocklength $n$.

A rate $R$ is *achievable* if for all $\delta > 0$, there exists a valid strategy and $n_0\left(\delta, \epsilon_0, \{\epsilon_{x,i}\}_{i=1}^{\infty}, R_{\mathrm{p}}\right)$ such that for the strategy's codes with blocklength $n \geq n_0$,

$$\mathbb{P}(\hat{W}^n_{\lfloor n(R-\delta) \rfloor} \neq W_{\lfloor n(R-\delta) \rfloor}) \leq \delta \ . \tag{13}$$

The set of achievable $R$ is denoted as $\mathcal{R}(\epsilon_0, \{\epsilon_{x,i}\}_{i=1}^{\infty}, R_{\mathrm{p}})$.

The *rate-interference budget (RIB) function* $R_{\text{IB}}(\epsilon_0, \{\epsilon_{x,i}\}_{i=1}^{\infty}, R_{\text{p}})$ is defined as

$$R_{\text{IB}}\left(\epsilon_0, \{\epsilon_{x,i}\}_{i=1}^{\infty}, R_{\text{p}}\right) = \sup_{R \in \mathcal{R}(\epsilon_0, \{\epsilon_{x,i}\}_{i=1}^{\infty}, R_{\text{p}})} R. \tag{14}$$

For the special case in which $\epsilon_{x,i} = \epsilon_x$ for all $i$, it will be convenient to use the shorthand $\vec{\epsilon}$, where $\vec{\epsilon}$ is a length $|\mathcal{X}|$ vector, and we will use the shorthand $R_{\text{IB}}(\vec{\epsilon}, R_{\text{p}})$.

## C. Contributions

We now state the main contributions of this paper. First, we find a valid strategy that achieves positive rates for the cognitive radio. From the definition of a valid strategy, this implies that there exists a sequence of codes such that the primary meets its target rate irrespective of $\epsilon_0, \{\epsilon_{x,i}\}_{i=1}^{\infty}$.

*Proposition 1:* For all $\nu > 0$ and corresponding $\epsilon_0, \{\epsilon_{x,i}\}_{i=1}^{\infty}, R_{\text{p}}$, a lower-bound to the RIB function is at least

$$R_{\text{IB}}\left(\epsilon_0, \{\epsilon_{x,i}\}_{i=1}^{\infty}, R_{\text{p}}\right) \geq (1 - R_{\text{p}}/(1 - \epsilon_0)) \cdot C^* , \tag{15}$$

where $C^* = \max_{p(x)} I(X;Y)$. Moreover, there exists a valid strategy that achieves the above rates for all $(\epsilon_0, \{\epsilon_{x,i}\}_{i=1}^{\infty}, R_{\text{p}})$ satisfying $\nu > 0$.

Proposition 1 follows immediately from Theorem 1. Furthermore, we can precisely characterize the capacity of the cognitive radio for the case of time-invariant interference on the primary, in which $\epsilon_{x,i} = \epsilon_x$ for all $x \in \mathcal{X}$.

*Proposition 2:* For all $\nu > 0$ and cases in which $\epsilon_{x,i} = \epsilon_x$ for all $i$, the RIB function is

$$R_{\text{IB}}(\vec{\epsilon}, R_{\text{p}}) = \max_{\substack{p(x): \\ \sum_x \epsilon_x p(x) \leq 1 - R_{\text{p}}}} I(X;Y) . \tag{16}$$

Moreover, there exists a valid strategy that achieves the above rates for all $(\vec{\epsilon}, R_{\text{p}})$ satisfying $\nu > 0$.

For this setting, we will refer to the constraint $\sum_x \epsilon_x p(x) \leq 1 - R_{\text{p}}$ as the *interference budget*. Note that the constraint is based on how much interference each of the cognitive radio's channel inputs generates on the primary compared to how much is tolerable for the primary's desired performance.

Proposition 2 follows immediately from Theorem 2, which provides achievability, and Theorem 3, which provides the converse. These are stated in Sections III and IV, respectively. We note that Theorem 2 relies on a more intricate valid strategy than the one in the proof of Theorem 1.

## III. ACHIEVABLE STRATEGIES

In this section, we present two achievable strategies and state results on the rates the cognitive radio can achieve while guaranteeing rate to the primary under various interference conditions. The first of these– the fixed-codebook protocol– is a generalization of the approach considered in Example 1, in which the cognitive radio becomes active only when the primary is meeting its target rate. We show that this strategy is valid, i.e. the primary meets its rate target under unknown time-varying interference characteristics, and can give equally general rate guarantees for the cognitive radio. The second strategy– the codebook-adaptive protocol– builds on the first strategy to predict the amount of interference the cognitive radio will generate on the primary and optimize its codebook to maximize its own rate. Like the first strategy, this strategy is also valid, so the primary meets its rate target under unknown time-varying interference characteristics. We provide rate guarantees for the cognitive radio under the more limited set of unknown time-invariant interference characteristics, and in Section IV, we show that the codebook-adaptive protocol provides the optimum rate for the cognitive radio within this set.

### A. Fixed-Codebook Protocol

Recall the approach considered in Example 1 over the noiseless channel. The silent symbol $x_{\text{off}}$ is used for each channel use when the primary is not meeting its target rate. Otherwise, one of the remaining $P$ symbols is used to send information about the message. As demonstrated in that example, this leads to a rate proportional to $\log P$. However, this strategy appears to be wasteful in that $x_{\text{off}}$ is not being used to send information about the message.

One way to overcome this limitation is to group multiple channel uses into *frames*. Each frame is either silent – consisting of only silent symbols $x_{\text{off}}$ – or active – consisting of any combinations of *all* $P + 1$ symbols, including $x_{\text{off}}$. Clearly, over the active frames, this increases the rate since the available channel input alphabet is larger. The main issues are:

- To find a rule by which the cognitive transmitter decides before each frame whether the frame will be silent or active. The cognitive transmitter then also needs some way of indicating its choice to the cognitive receiver.

- To appropriately select the frame length. If the frame length is too short, then no rate gain is attained. Conversely, if the frame length is too large, then the non-interference guarantee given in (12) can no longer be respected.

We now illustrate the approach in the context of Example 1. For the sake of concreteness, consider the case in which the frame length $K_n = 3$ channel uses. For this illustration, we will assume the decision to become active is governed by the threshold rule $\sum_{j=1}^{3\lfloor (i-1)/3 \rfloor}\left(A_j - \frac{1}{2}\right) > 0$. Then a sample run may look as follows:

| $i$ | **1** | 2 | 3 | **4** | 5 | 6 | **7** | $\cdots$ |
|---|---|---|---|---|---|---|---|---|
| $\sum_{j=1}^{i-1}\left(A_j - \frac{1}{2}\right)$ | **0** | $\frac{1}{2}$ | 1 | $\mathbf{\frac{1}{2}}$ | 0 | $-\frac{1}{2}$ | $-1$ | $\cdots$ |
| $X_i$ | $\mathbf{x_{\text{off}}}$ | $x_{\text{off}}$ | $x_{\text{off}}$ | $\mathbf{x_{\text{on,1}}}$ | $x_{\text{off}}$ | $x_{\text{on,P}}$ | $x_{\text{off}}$ | $\cdots$ |

Times $i = 4, 5, 6$ represent an active frame, where the channel input at time $i = 4$ is simply a beacon to indicate to the decoder that the frame is active; the message information is sent over $i = 5, 6$. Despite the fact that the primary meets the rate target $R_{\text{p}} = \frac{1}{2}$ over channel uses 2 and 3, the cognitive radio sends the silent symbol $x_{\text{off}}$ for the duration of the frame. Thus, one has to be careful to set the frame length $K_n$ and transmission threshold to make sure the cognitive radio can achieve a significant rate. Likewise, the cognitive radio sends the message information $(x_{\text{off}}, x_{\text{on,P}})$ over channel uses 5 and 6 even though the primary no longer exceeds the rate target $\frac{1}{2}$. Thus, one has to be careful to set the frame length $K_n$ and transmission threshold so that the primary's rate satisfies (12), so the strategy is valid.

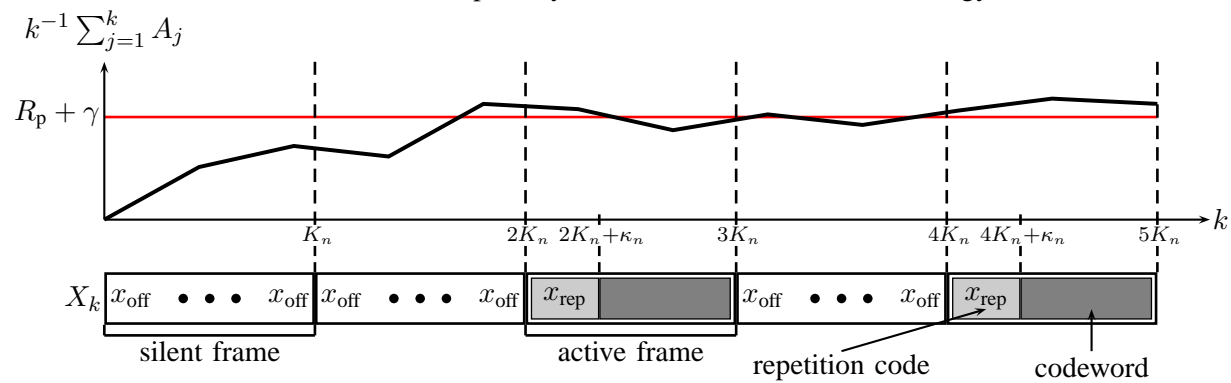

Fig. 3. In the fixed-codebook protocol, channel uses are grouped into units known as frames. At the start of a frame, the cognitive radio encoder chooses to become active if the primary's packet rate $k^{-1}\sum_{j=1}^{k} A_j$ is above a threshold $R_{\text{p}} + \gamma + o(1)$. Otherwise, it stays silent for the frame, i.e. sends the symbol $x_{\text{off}}$. On an active frame, the encoder uses a length $\kappa_n$ repetition code to signal to the decoder that it is active and sends a codeword over the remaining channel uses to convey additional bits of the message.

We now use the intuition from the illustration above to construct the *fixed-codebook protocol*, which we will then prove is a valid strategy, as defined in Section II. Figure 3 provides an illustration of the fixed-codebook protocol. For convenience, we define

$$S_k = \sum_{i=1}^{k} (A_i - R_{\text{p}}) \ , \tag{17}$$

which is positive at time $k$ if and only if the primary is exceeding its target rate.

*1) Determining Silent Frames:* As before, the cognitive radio makes a decision to be silent or active over frames of length $K_n$ channel uses. Specifically, the following condition specifies the frames over which the cognitive radio is silent:

$$X_j = x_{\text{off}} \quad \text{if} \quad S_{i_j} - i_j \gamma < K_n \ , \tag{18}$$

where $i_j = \lfloor (j-1)/K_n \rfloor \cdot K_n$, and $\gamma$ is an additional parameter for setting the threshold along with $K_n$ to satisfy condition (12).

*2) Active Frames:* It remains to define what the cognitive transmitter does over an active frame. As in the noiseless case, we want to inform the decoder that the frame is active, but in the noisy case, it cannot be done with a single channel use.

*a) Repetition Coding:* The cognitive transmitter uses a length $\kappa_n$ repetition code to inform the cognitive receiver that the frame is active. We will assume without loss of generality there exists a channel input $x_{\text{rep}} \neq x_{\text{off}}$ such that $\mathbb{P}(Y = y | X = x_{\text{off}}) \neq \mathbb{P}(Y = y | X = x_{\text{rep}})$ for some $y$. (Note: we can assume this without loss of generality since if it were not true for any symbol, the cognitive radio's channel inputs $X_i$ would be independent of the channel outputs $Y_i$, and the channel could not be used for communicating in the first place.) Then the repetition code over the first $\kappa_n$ channel uses of an active frame is specified by the following condition:

$$X_j = x_{\text{rep}} \quad \text{if} \quad S_{i_j} - i_j \gamma \geq K_n, \ i_j < j \leq i_j + \kappa_n \ , \tag{19}$$

where $i_j = \lfloor (j-1)/K_n \rfloor \cdot K_n$.

*b) Message Information:* For the remaining channel uses of an active frame, the encoder sends information about the message to the decoder. It does so with a blocklength $K_n - \kappa_n$ codebook $\mathcal{C}_{\text{fixed}}$ of rate $C^* - \tilde{\delta}$, where $C^* = \max_{p(x)} I(X;Y)$. We will denote codeword $m$ as $\tilde{X}^{K_n - \kappa_n}(m)$, where $m \in \{1, \ldots, \exp\{(K_n - \kappa_n)(C^* - \tilde{\delta})\}\}$.

The following notation will be useful for understanding the channel inputs during the remainder of an active frame. Let $V_1$ denote the channel index preceding the start of the first active frame, $V_2$ the second, $V_3$ the the third, and so on. That is,

$$V_1 = \inf\{i \geq 0 : S_i - i\gamma \geq K_n \ , i = mK_n \text{ for some } m \in \mathbb{Z}\} \ , \tag{20}$$

$$V_k = \inf\{i > V_{k-1} : S_i - i\gamma \geq K_n \ , i = mK_n \text{ for some } m \in \mathbb{Z}\} \ . \tag{21}$$

We now characterize the remaining channel inputs. For the $\ell$-th active frame and letting $m_\ell$ be bits $\ell(K_n - \kappa_n)(C^* - \tilde{\delta}) \log_2 e + 1$ through $(\ell+1)(K_n - \kappa_n)(C^* - \tilde{\delta}) \log_2 e$ of message $W$,

$$X_j = \tilde{X}_{j - V_\ell - \kappa_n}(m_\ell) \quad \text{if} \quad V_\ell + K_n \geq j > V_\ell + \kappa_n \ . \tag{22}$$

 

A summary of the fixed-codebook protocol is given in Table I.

| $X_j$ | Conditions | Description |
|---|---|---|
| $\tilde{X}_{j-V_\ell - \kappa_n}(m_\ell)$ | $V_\ell + K_n \geq j > V_\ell + \kappa_n, \ell \in \mathbb{Z}^+$ | Active frame: $\tilde{X}^{K_n - \kappa_n}(m_\ell) \in \mathcal{C}_{\text{fixed}}$ to send fragment $m_\ell$. |
| $x_{\text{rep}}$ | $V_\ell + \kappa_n \geq j > V_\ell, \ell \in \mathbb{Z}^+$ | Repetition code: notify decoder of active frame |
| $x_{\text{off}}$ | all other $j$ | Silent frame |

*3) Performance of the Fixed-Codebook Protocol:* For this strategy, we have the following result.

*Theorem 1:* For all $R_{\text{p}}, \nu > 0$, there exist choices of $\kappa_n, K_n, \gamma, \tilde{\delta}$ such that the fixed-codebook protocol is a valid strategy, i.e. primary's packet rate satisfies the condition in (12) for all $\{\epsilon_{x,i}\}_{i=1}^{\infty}$, $x \in \mathcal{X} - \{x_{\text{off}}\}$. Furthermore, for these parameter choices, the rate

$$\left(1 - \frac{R_{\text{p}}}{1 - \epsilon_0}\right) \cdot C^* \tag{23}$$

is achievable for the cognitive radio, where $C^* = \max_{p(x)} I(X;Y)$.

*Proof:* While other choices will work, for the purposes of the proof, we will let

$$K_n = \lfloor n^{1/8} \rfloor, \kappa_n = \lfloor n^{1/16} \rfloor, \tag{24}$$

and any $\gamma$ satisfying $0 < \gamma < \min\{\tilde{\delta}/2, \nu/2\}$. We will assume that $\tilde{\delta} > 0$, but a detailed prescription is given in Lemma 6 below to allow the rate loss to become arbitrarily small.

The proof of the theorem is divided into three parts.

1) The primary's rate satisfies condition (12), so the strategy is valid. (Lemma 3)

2) There exists a codebook such that cognitive radio decoder error probability is small, thus satisfying (13) for some $R$. (Lemma 4)

3) By appropriately choosing $\tilde{\delta}$, the $R$ in (13) can be made arbitrarily close to $\left(1 - \frac{R_{\text{p}}}{1 - \epsilon_0}\right) \cdot C^*$. (Lemma 6)

These results are proved in the Appendix I. ■

## B. Codebook-Adaptive Protocol

Let us return to Example 1. Theorem 1 implies that when the fixed-codebook protocol is applied to the noiseless channel, the cognitive radio is guaranteed to achieve rates

$$R \geq (1 - R_{\text{p}}/(1 - \epsilon_0)) \cdot \log(1 + P). \tag{25}$$

Hence, the fixed-codebook protocol only adapts the duty-cycle to the actual degree of interference. In equation (25), when $P$ is large, this is not a good strategy. A better strategy would be to also adapt the *codebook* to the actual degree of interference, thereby guaranteeing a higher duty-cycle. Clearly, to still meet the interference guarantee the rate of the adapted codebook will typically be smaller, but with respect to equation (25), this penalty will appear in the logarithm. In this section, we propose the *codebook-adaptive protocol*, which first obtains a coarse measurement of the actual interference (Phase I), uses this to select a codebook appropriately and communicates its choice to the decoder (Phase II), and then runs the standard fixed-codebook protocol described in Section III-A (Phase III).

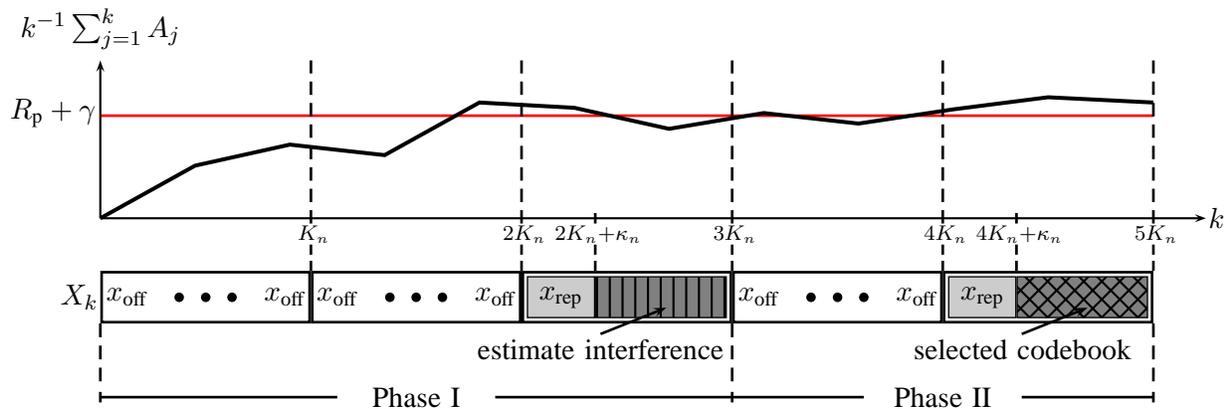

Fig. 4. The codebook-adaptive protocol is like the fixed-codebook protocol except the first two active frames are used to select a codebook to use and inform the decoder about it. In Phase I, the cognitive radio sends pilots of each of its channel inputs and uses the ARQs to create estimates of the interference it generates on the primary. In Phase II, the cognitive radio notifies the decoder which among a polynomial sized set of codebooks it has selected based on its estimates from Phase I. Phase III, which immediately follows the end of Phase II above, is almost identical to the fixed-codebook protocol, except the codewords are now from the codebook selected during Phase I and Phase II.

The codebook-adaptive protocol is summarized in Figure 4. The strategy is quite similar to the fixed-codebook protocol. In fact, it uses the same threshold rule and the same repetition code to signify an active frame. That is, the codebook-adaptive protocol follows the rules:

$$X_j = x_{\text{off}} \quad \text{if} \quad S_{i_j} - i_j \gamma < K_n \ , \tag{26}$$

$$X_j = x_{\text{rep}} \quad \text{if} \quad S_{i_j} - i_j \gamma \ge K_n, \ i_j < j \le i_j + \kappa_n \ , \tag{27}$$

where $i_j = \lfloor (j-1)/K_n \rfloor \cdot K_n$, are identical to conditions (18) and (19) in the fixed-rate protocol.

The difference between the two strategies is thus in what follows the repetition code in an active frame. In particular, the encoder uses the first active frame to estimate the channel, the second to inform the decoder which codebook it will use based on those rates, and the third and greater active frames to send message information using the selected codebook.

As before, let $V_1$ denote the channel index preceding the start of the first active frame, $V_2$ the second, $V_3$ the the third, and so on. That is, $V_1 = \inf\{i \geq 0 : S_i - i\gamma \geq K_n , i = mK_n$ for some $m \in \mathbb{Z}\}$ and $V_k = \inf\{i > V_{k-1} : S_i - i\gamma \geq K_n , i = mK_n$ for some $m \in \mathbb{Z}\}$.

*1) Phase I:* During the first active frame, the cognitive radio estimates the interference produced by each channel input. Let $\mu = \lfloor \frac{K_n - \kappa_n}{|\mathcal{X}|} \rfloor$. Then for $x \in \{0, \dots, |\mathcal{X}| - 1\}$, the channel inputs for the first frame can be described as

$$X_j = x , \quad \text{if} \quad V_1 + \kappa_n + (x+1)\mu \geq j > V_1 + \kappa_n + x\mu . \tag{28}$$

Using these channel inputs, the encoder can use the ARQs to estimate the primary's erasure probabilities.

$$\hat{\epsilon}_x = \mu^{-1} \sum_{i=V_1+\kappa_n+x\mu+1}^{V_1+\kappa_n+(x+1)\mu} A_i . \tag{29}$$

With these estimates, the end of this first active frame marks the end of Phase I.

*2) Phase II :* Based on the estimates $\hat{\epsilon}_x$, the encoder chooses a codebook among a set of codebooks; it informs the decoder of this choice in Phase II.

Each codebook in the set has a different input distribution corresponding uniquely to each length-$C_n$ type $p_{x^{C_n}}$ of $\mathcal{X}$, i.e. $p_{x^{C_n}}$ is a probability distribution with the property that for all $x \in \mathcal{X}$, $p_{x^{C_n}}(x) = n_x/C_n$ such that $n_x$ is a nonnegative integer and $\sum_{x \in \mathcal{X}} n_x = C_n$. Thus, there are at most $(C_n + 1)^{|\mathcal{X}|}$ codebooks in the set. The codebook $\mathcal{C}_{x^{C_n}}$ of type $p_{x^{C_n}}$ is a random codebook with codewords generated i.i.d. according to $\prod_{k=1}^{K_n - \kappa_n} p_{x^{C_n}}(x_k)$ and has

$$\tilde{M}_{x^{C_n}} = \exp\{(K_n - \kappa_n)(R_{x^{C_n}} - \tilde{\delta})^+\} , \tag{30}$$

where $R_{x^{C_n}}$ is the mutual information $I(X;Y)$ with $X$ having input probability distribution $p_{x^{C_n}}(x)$. One then selects the codebook according to the following rule:

$$\chi = \underset{\substack{x^{C_n}: \\ \sum_x \hat{\epsilon}_x p_{x^{C_n}}(x) \leq 1 - R_p - 2\gamma - \tilde{\delta}}}{\operatorname{argmax}} \tilde{M}_{x^{C_n}} . \tag{31}$$

The encoder uses the codebook from the fixed-codebook protocol in the second active frame to inform the decoder of its codebook selected codebook. (Note: Based on the parameter choices considered in this work, $(C_n + 1)^{|\mathcal{X}|}$ is small enough to only require a messages from fixed-codebook protocol's codebook, so the encoder simply uses the messages that result in the lowest probability of error.) Suppose the selected codebook corresponds to message $m^\chi$. Then for the second active frame,

$$X_j = \tilde{X}_{j-V_2-\kappa_n}(m^\chi) \quad \text{if} \quad V_2 + K_n \geq j > V_2 + \kappa_n \ . \tag{32}$$

With the decoder informed of which codebook has been selected, the end of this active frame marks the end of Phase II.

*3) Phase III:* In Phase III, the active frames are now used to send message information. Thus, they resemble the active frames in the fixed-codebook protocol, with the main difference that the codebook $\chi$ is used.

Let $m_\ell$ be bits $\ell(K_n - \kappa_n)(R_\chi - \tilde{\delta})\log_2 e + 1$ through $(\ell + 1)(K_n - \kappa_n)(R_\chi - \tilde{\delta})\log_2 e$ of message $W$. For the $(\ell + 2)$-th active frame, we can express the message information segment of the frame as

$$X_j = \tilde{X}_{j-V_{\ell+2}-\kappa_n}(m_\ell) \quad \text{if} \quad V_{\ell+2} + K_n \geq j > V_{\ell+2} + \kappa_n \ , \tag{33}$$

where $\tilde{X}^{K_n - \kappa_n}(m_\ell) \in \mathcal{C}_\chi$.

A summary of the codebook-adaptive protocol is given in Table II.

TABLE II

SUMMARY OF THE CODEBOOK-ADAPTIVE PROTOCOL.

| $X_j$ | Conditions | Description |
|---|---|---|
| $x$ | $V_1 + \kappa_n + (x+1)\mu \geq j > V_1 + \kappa_n + x\mu$ | Phase I: estimate $x$'s interference with primary's ARQs. |
| $\tilde{X}_{j-V_\ell-\kappa_n}(m^\chi)$ | $V_2 + K_n \geq j > V_2 + \kappa_n$ | Phase II: $\tilde{X}^{K_n - \kappa_n}(m^\chi) \in \mathcal{C}_{\text{fixed}}$ for selection $\mathcal{C}_\chi$. |
| $\tilde{X}_{j-V_\ell-\kappa_n}(m_\ell)$ | $V_{\ell+2} + K_n \geq j > V_{\ell+2} + \kappa_n, \ell \in \mathbb{Z}^+$ | Phase III: $\tilde{X}^{K_n - \kappa_n}(m_\ell) \in \mathcal{C}_\chi$ to send fragment $m_\ell$. |
| $x_{\text{rep}}$ | $V_\ell + \kappa_n \geq j > V_\ell, \ell \in \mathbb{Z}^+$ | Repetition code: notify decoder of active frame |
| $x_{\text{off}}$ | all other $j$ | Silent frame |

We now state the result for the codebook-adaptive protocol.

*Theorem 2:* For all $R_{\text{p}}, \nu > 0$, there exists a choice of $K_n, \kappa_n, C_n, \gamma, \tilde{\delta}$ such that the codebook-adaptive protocol is a valid strategy, i.e. the primary's packet rate satisfies the condition in (12) for all $\{\epsilon_{x,i}\}_{i=1}^\infty$, $x \in \mathcal{X} - \{x_{\text{off}}\}$. Furthermore, when the interference on the primary is time-invariant, i.e. $\epsilon_{x,i} = \epsilon_x$ for $x \in \mathcal{X} - \{x_{\text{off}}\}$, the rate

$$\max_{\substack{p(x): \\ \sum_x \epsilon_x p(x) \leq 1 - R_{\text{p}}}} I(X;Y) \tag{34}$$

is achievable for the cognitive radio under the same parameter settings.

*Proof:* The parameters $K_n, \kappa_n$ will be set as in (24), $C_n = \sqrt{\kappa_n}$, and any $\gamma$ satisfying $0 < \gamma < \min\{\tilde{\delta}/2, \nu/2\}$. We will assume $\tilde{\delta} > 0$, but a detailed prescription is given in Lemma 12 below to get arbitrarily close to the rate in the statement of the theorem.

The proof of the theorem is divided into three parts.

1) As in the fixed-rate protocol, we can apply Lemma 3 since (18) and (26) are identical conditions. Thus, the primary's rate satisfies the condition (12), so the strategy is valid.

2) The cognitive radio decoder error probability is small, thus satisfying (13) for some $R$. (Lemma 7)

3) By appropriately choosing $\tilde{\delta}$, the $R$ in (13) can be made arbitrarily close to $R_{\text{IB}}(\vec{\epsilon}, R_{\text{p}})$ with probability going to 1 as $n \to \infty$. (Lemma 12)

With the exception of Lemma 3, these results are proved in the Appendix II. ∎

## IV. Converse

To show the converse, we will relax the conditions stipulated in the problem setup, thereby allowing a larger class of strategies. It turns out that in some cases, this larger class does not increase the rate region.

*Theorem 3:* For all $\nu > 0$ and for $\epsilon_{x,i} = \epsilon_x$ for all $i$,

$$R_{\text{IB}}(\vec{\epsilon}, R_{\text{p}}) \leq \max_{\substack{p(x): \\ \sum_x \epsilon_x p(x) \leq 1 - R_{\text{p}}}} I(X; Y) \tag{35}$$

*Proof:* From the definition of achievable rate,

$$nR \leq H(W_{\lfloor n(R-\delta) \rfloor}) - n\delta + 1 \tag{36}$$

$$\leq I(W_{\lfloor n(R-\delta) \rfloor}; Y^n) + 2n\delta + 1 \tag{37}$$

$$= \sum_{i=1}^{n} H(Y_i | Y^{i-1}) - H(Y_i | Y^{i-1}, W_{\lfloor n(R-\delta) \rfloor}) + 2n\delta + 1 \tag{38}$$

$$\leq \sum_{i=1}^{n} H(Y_i) - H(Y_i | Y^{i-1}, W, A^{i-1}, X^i) + 2n\delta + 1 \tag{39}$$

$$= \sum_{i=1}^{n} H(Y_i) - H(Y_i | X_i) + 2n\delta + 1 \tag{40}$$

$$= \sum_{i=1}^{n} I(X_i; Y_i) + 2n\delta + 1 \;, \tag{41}$$

where (37) follows from Fano's inequality, (38) from the chain rule, (39) since conditioning cannot increase entropy, (40) by the Markov chain $(W, A^{i-1}, Y^{i-1}, X^{i-1}) \leftrightarrow X_i \leftrightarrow Y_i$, and (41) by definition.

We have yet to place a restriction on the strategies. Recall that valid strategies need to satisfy the condition in (12). If condition (12) is satisfied, then the code of blocklength $n$ satisfies

$$n^{-1} \sum_{i=1}^{n} \mathbb{E}\left[A_i\right] \geq R_{\mathsf{p}} - K_{1, R_{\mathsf{p}}, \nu, n} e^{-n \cdot K_{2, R_{\mathsf{p}}, \nu}} \ . \tag{42}$$

We now consider only this weaker condition on the channel inputs as opposed to the stronger one given by (12). By the concavity of mutual information with respect to its input distribution, we can combine (41) and (42) to yield that for all $\delta > 0$, there exists large enough $n$ such that

$$R \leq \max_{\substack{p(x): \\ \sum_x \epsilon_x p(x) \leq 1 - R_{\mathsf{p}} + \delta}} I(X; Y) + 3\delta \tag{43}$$

Since $\delta$ can be made arbitrarily small, we can conclude the result. ∎

## V. Examples

Propositions 1 and 2 provide a lower bound and an exact result for the RIB function under different interference conditions, respectively. In this section, we evaluate the RIB function given in Proposition 2 for cases in which the interference characteristics on the primary are time-invariant. We then evaluate the RIB function lower bound given in Proposition 1 for these examples when the interference characteristics are time-varying.

### A. Evaluation of the RIB Function for Time-Invariant Interference Characteristics

We first explore the setting in which the interference parameters $\epsilon_{x,i}$ are time-invariant, i.e. $\epsilon_{x,i} = \epsilon_x$ for all $i, x$. In this setting, Proposition 2 gives an exact expression for the RIB function $R_{\mathsf{IB}}(\vec{\epsilon}, R_{\mathsf{p}})$.

We first evaluate the RIB function for Example 1. We first rewrite the expression in Proposition 2 as

$$R_{\mathsf{IB}}(\vec{\epsilon}, R_{\mathsf{p}}) = \max_{\substack{p(x): \\ \sum_x \epsilon_x p(x) \leq 1 - R_{\mathsf{p}}}} I(X; Y) \tag{44}$$

$$= \max_{\substack{p(x): \\ p(x \neq x_{\mathsf{off}})\epsilon_1 + p(x = x_{\mathsf{off}})\epsilon_0 \leq 1 - R_{\mathsf{p}}}} H(X)$$

$$= \max_{p \leq \frac{1 - R_{\mathsf{p}} - \epsilon_0}{\epsilon_1 - \epsilon_0}} h(p) + p \log P$$

$$= \begin{cases} \log(P + 1) \ , & b \geq \frac{P}{P+1} \\ h(b) + b \log P \ , & \text{otherwise} \end{cases} , \tag{45}$$

where $b = \frac{1 - R_{\mathsf{p}} - \epsilon_0}{\epsilon_1 - \epsilon_0}$. Figure 5 shows (45) in terms of $b$, which we can think of as a summary of the interference budget.

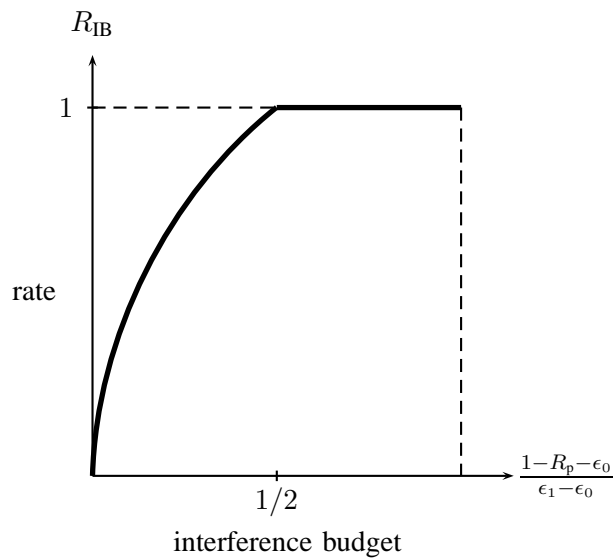

Fig. 5. A schematic plot of the RIB function for Example 1 when $P = 1$.

*Example 2:* Consider a DMC with $|\mathcal{X}| = 1 + P$ channel input symbols and $|\mathcal{Y}| = P$ output symbols with the following property:

$$\mathbb{P}(Y = y | X = x) = \begin{cases} 1, & y = x, \quad x \in \{1, \ldots, P\} \\ \frac{1}{P}, & y \in \mathcal{Y}, \quad x = x_{\text{off}} \end{cases}. \tag{46}$$

If $\epsilon_0 = 0$, $\epsilon_x = 1$ for $x \in \{1, \ldots, P\}$, then evaluating the RIB function from Proposition 2 yields

$$R_{\text{IB}}(\vec{\epsilon}, R_{\text{p}}) = (1 - R_{\text{p}}) \log_2 P, \tag{47}$$

where the units are in bits per channel use.

We now consider a case in which the secondary has an alternative to $x_{\text{off}}$ to control interference. The channel model resembles the one in Example 2, except there are now additional channel inputs.

*Example 3:* Let $P$ be even and consider a DMC with $|\mathcal{X}| = 1 + 3P/2$ channel input symbols and $|\mathcal{Y}| = P$ output symbols with the following property:

$$\mathbb{P}(Y = y | X = x) = \begin{cases} 1, & y = x, \quad x \in \{1, \ldots, P\} \\ \frac{1}{P}, & y \in \mathcal{Y}, \quad x = x_{\text{off}} \\ \frac{1}{2}, & y = 2(x - P) - 1, \quad x \in \{P + 1, \ldots, P + P/2\} \\ \frac{1}{2}, & y = 2(x - P), \quad x \in \{P + 1, \ldots, P + P/2\} \end{cases}. \tag{48}$$

An illustration of these transition probabilities are given in Figure 6. We now consider the case in which $\epsilon_0 = 0$, $0 < R_{\text{p}} < 1$, $\epsilon_x = 1$ for $x \in \{1, \ldots, P\}$, and $\epsilon_x = \epsilon_{1/2} < 1 - R_{\text{p}}$ for $x \in \{P + 1, \ldots, P + P/2\}$.

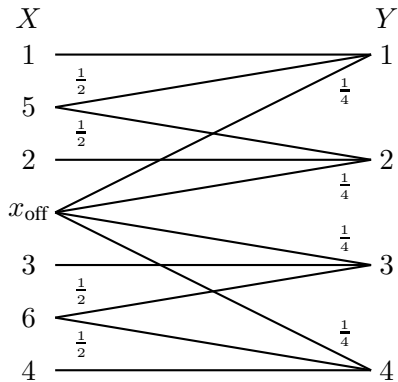

Fig. 6. Illustration of transition probabilities for Example 3 when $P = 4$.

Under these assumptions, evaluating the RIB function from Proposition 2 yields

$$R_{\mathrm{IB}}(\vec{\epsilon}, R_{\mathrm{p}}) = \max_{p(x):\sum_x \epsilon_x p(x) \leq 1 - R_{\mathrm{p}}} I(X;Y) \tag{49}$$

$$= \frac{1 - R_{\mathrm{p}} - \epsilon_{1/2}}{1 - \epsilon_{1/2}} \log_2 P + \frac{R_{\mathrm{p}}}{1 - \epsilon_{1/2}} \log_2(P/2) \tag{50}$$

$$= \log_2 P - \frac{R_{\mathrm{p}}}{1 - \epsilon_{1/2}} , \tag{51}$$

where the units are in bits per channel use. Note that the rate loss due to the primary can be at most 1 bit in this setting. Moreover, this can be arbitrarily better than the case in Example 2 by making $P$ large and $R_{\mathrm{p}}$ close to 1, for which the target rate $R_{\mathrm{p}}$ induced a multiplicative penalty on the $\log_2 P$ term in (47).

## B. Further Considerations for Time-Varying Interference Characteristics

The most interesting and realistic scenarios concern the case when the interference characteristics are time-varying. The codebook-adaptive protocol introduced in Section III can deal with this as long as it is well behaved. However, for some "maliciously chosen" time-varying characteristics, the proposed startegy can be fooled into choosing a low rate codebook in Phase II when the interference conditions are less severe in Phase III. The effect of such a possibility is illustrated in Figure 7. One option might be to consider a strategy that periodically readapts the codebook, which, while potentially beneficial, is outside the scope of this work. Instead, we consider the simpler strategy given by the fixed-codebook protocol.

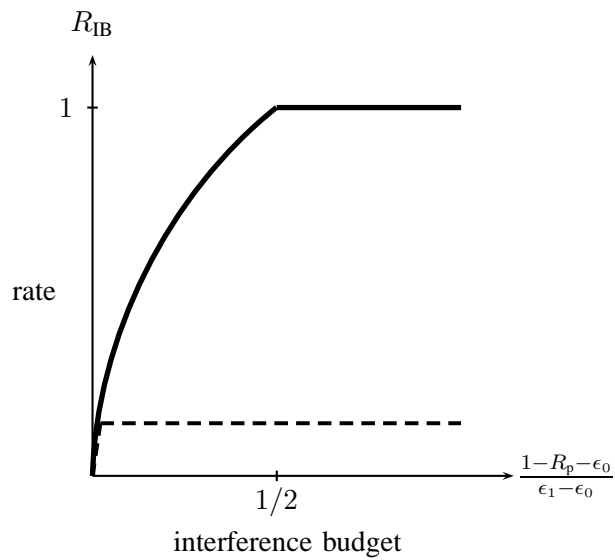

Fig. 7. A schematic plot of the RIB function for Example 1 when $P = 1$. The dashed line suggests how maliciously chosen time-varying characteristics can cause the encoder to select a "low rate" codebook, which saturates well below the actual RIB function when the interference budget is large.

For Example 1, Proposition 1 implies the fixed-codebook protocol lets the cognitive radio achieve the rate

$$R_{\text{IB}}\left(\epsilon_0, \{\epsilon_{x,i}\}_{i=1}^{\infty}, R_{\text{p}}\right) \geq \left(1 - R_{\text{p}}/(1 - \epsilon_0)\right) \log(1 + P) \tag{52}$$

for all $\{\epsilon_{x,i}\}_{i=1}^{\infty}$, $x \neq x_{\text{off}}$. For the restricted time-invariant interference setting of Example 1, can use $b = \frac{1 - R_{\text{p}} - \epsilon_0}{\epsilon_1 - \epsilon_0}$ to compare its performance against the RIB function. It turns out that for $b > b^*$, the codebook chosen by the codebook-adaptive protocol has the same asymptotic rate and produces the same interference on the primary as that in the fixed-codebook protocol. Thus, depending on one's assumptions about the interference environment, there are instances in which the fixed-codebook protocol may be preferable to the codebook-adaptive protocol.

Despite these guarantees, there are situations in which the fixed-codebook protocol can be arbitrarily worse. Recall Examples 2 and 3. It turns out that in both cases when $\epsilon_0 = 0$, Proposition 1 implies the fixed-codebook protocol guarantees rates given by

$$R_{\text{IB}}\left(0, \{\epsilon_{x,i}\}_{i=1}^{\infty}, R_{\text{p}}\right) \geq \left(1 - R_{\text{p}}\right) \log_2 P \ , \tag{53}$$

which matches the RIB function in (47) for Example 2. However, as already illustrated, by making $P$ large and $R_{\text{p}}$ close to 1, the RIB function in Example 3, given in (51), can be made arbitrarily larger than the one in Example 2. This implies that the loss for applying the fixed-codebook protocol can be significant. Thus, one's choice between these two protocols depends jointly on the cognitive radio's channel and the interference generated on the primary. Indeed, there may exist strategies that can trade off the competing desires of optimality and robustness better than the ones proposed. These are discussed further in the next section.

## VI. DISCUSSION

In this paper, a novel model was proposed for a cognitive radio problem. The basic problem is that the cognitive radio must not disturb the primary user (i.e., the license holder). The specific aspect of our model is that the cognitive radio is ignorant of the channel characteristics according to which it interferes with the primary. To mitigate this uncertainty, the cognitive radio may eavesdrop on the primary system's ARQ feedback signal. We show how this can be exploited to design two adaptive cognitive radio strategies, each of which provides a fixed rate guarantee to the primary and variable rate guarantee to the cognitive radio that depends on its interference budget, the amount of interference it is allowed to generate on the primary user. The problem statement and results provide a starting point for new research directions and problems, some of which we briefly outline in the sequel.

### A. Gaussian Channels

In this work, the cognitive radio's channel is a DMC with each symbol affecting the primary's erasure probability. An analogous model and result for the Gaussian setting would be desirable to gain further intuitions about the design of a cognitive radio system. For instance, if the primary employs a Gaussian codebook that assumes a certain level of interference, the cognitive radio may use the ARQs to choose the highest power codebook that maintains that level of interference on the primary.

### B. Primary with a Fixed Delay Constraint

In our model, the cognitive radio must operate such that eventually, the primary attains its prespecfied target rate. A more restrictive setting would be to also enforce a delay constraint. That is, the cognitive radio must operate such as to not delay packets by more than a certain prespecified bound. Alternatively, this can be formulated as a "sliding window" rate constraint: over any window of a prespecified length, the primary must attain its prespecified rate. It would be interesting to understand by how much this lowers the "interference budget" of the cognitive radio, and thus, its capacity.

### C. Improved Strategies

The cognitive radio's rate guarantees for the fixed-codebook protocol are somewhat pessimistic, and the rate guarantees for the codebook-adaptive protocol are restricted to the smaller class of time-invariant interference parameters on the primary. The problem is that since the codebook-adaptive protocol only selects the codebook once, varying the interference conditions in the time can lead to suboptimal performance. For instance, the interference conditions in Phase I can be such that the primary selects a codebook with negligible rate in Phase II only to discover that there is no interference to the primary in Phase III. Thus, its performance can be significantly worse than the fixed-codebook protocol in the time-varying setting.

An obvious alternative would be a strategy that periodically readapts the codebook, which, if done properly, may be able to provide stronger rate guarantees than those already provided in the time-varying setting. One may also wish to restrict the set of codebooks so that all codebooks have a rate above a certain threshold. Then, arguments similar to those used for the fixed-codebook protocol can provide rate guarantees for the time-varying case, and one can also exploit the advantage afforded by adapting one's codebook for the time-invariant case.

### D. Multiple Cognitive Radios

In our model, there is only a single cognitive radio interfering with the primary. A more interesting situation will involve multiple cognitive radios all competing for the same interference budget. Clearly, this significantly changes the dynamics of the problem. Are there efficient strategies that give good rates for the cognitive radios while respecting the primary user? First of all, if all the cognitive radios have access to $A_k$ with different delays, then the arguments in this work would need to be extended. The existence of multiple users also leads to the issue that any individual cognitive radio may not cause significant interference to the primary by itself, but the aggregate interference from all cognitive radios can still be quite large. Another issue to consider is how the cognitive radios might divide their rate in an equitable way based not only on their own channels but also on how much interference each generates on the primary.

### E. Noisy feedback

In our model, the cognitive radio has a perfect observation of the ARQ signal of the primary, i.e., of the values of $A_k$. However, in practice there may be noise that corrupts the encoder's knowledge of $A_k$. This may also play a crucial role for the case of multiple cognitive radios, in which the noise may be different for different terminals in the system.

## Acknowledgments

We would like to acknowledge comments and suggestions from Anant Sahai and Sourav Chatterjee (Stat, UC Berkeley). Stimulating discussions with Mubaraq Mishra are also gratefully acknowledged. This work was supported by the NSF under award CNS-0326503.

## Appendix I

## Proof of Theorem 1

### A. Primary Meets Rate Target

*Lemma 1:* Let $S_k$ be defined as in (17). The sequence of random variables

$$M_k = e^{\lambda S_k - k(f_\lambda(\epsilon_0) - \lambda R_\mathfrak{p}) - \sum_{j=1}^k \tau_j(f_\lambda(\epsilon_{j,1}) - f_\lambda(\epsilon_0))} , \tag{54}$$

where $f_\lambda(\epsilon) = \log((1-\epsilon)e^\lambda + \epsilon)$ and $\tau_k = \mathbb{I}_{\{S_{\lfloor k/K_n \rfloor K_n} - K_n - \lfloor k/K_n \rfloor K_n \gamma \geq 0\}}$, forms a martingale.

*Proof:* First observe that we can express $M_k$ in terms of the recurrence equation

$$M_k = M_{k-1} e^{\lambda A_k - \tau_k(f_\lambda(\epsilon_{k,1}) - f_\lambda(\epsilon_0))} . \tag{55}$$

From this, we find that

$$\mathbb{E}[M_k | M_0, \ldots, M_{k-1}] = \mathbb{E}[M_k | S_0, \ldots, S_{k-1}] \tag{56}$$

$$= M_{k-1} e^{-\tau_k(f_\lambda(\epsilon_{k,1}) - f_\lambda(\epsilon_0))} \mathbb{E}[e^{\lambda A_k} | S_0, \ldots, S_{k-1}] \tag{57}$$

$$= M_{k-1} . \tag{58}$$

∎

*Lemma 2:* Let $S_k$ be defined as in (17) and $r$ be a positive integer and define the stopping time

$$N = r \cdot \inf\{i > 0 : S_{ir} - ir\gamma - r \geq 0\} . \tag{59}$$

Then for $R_\mathfrak{p} + \gamma \leq 1 - \epsilon_0$ and $s \leq 0$,

$$\mathbb{P}(N \geq t | S_0 = s) \leq \left(\frac{(1 - R_\mathfrak{p} - \gamma)(1 - \epsilon_0)}{(R_\mathfrak{p} + \gamma)\epsilon_0}\right)^{2r-s} e^{-tD(1 - R_\mathfrak{p} - \gamma \| \epsilon_0)} . \tag{60}$$

*Proof:* Consider the martingale (see Lemma 1)

$$M_k = e^{\lambda S_k - k(f_\lambda(\epsilon_0) - \lambda R_\mathfrak{p}) - \sum_{j=1}^k \tau_j(f_\lambda(\epsilon_{j,1}) - f_\lambda(\epsilon_0))} , \tag{61}$$

where $f_\lambda(\epsilon) = \log((1-\epsilon)e^\lambda + \epsilon)$ and $\tau_k = \mathbb{I}_{\{S_{\lfloor k/K_n \rfloor K_n} - K_n - \lfloor k/K_n \rfloor K_n \gamma \geq 0\}}$. The optional stopping theorem [21, Thm. 4.7.4, p. 270] implies

$$e^{\lambda s} = \mathbb{E}[M_{N \wedge m} | S_0 = s] , \tag{62}$$

where $N \wedge m$ denotes their minimum. We can substitute $\lambda = \log \frac{(R_{\mathrm{p}}+\gamma)\epsilon_0}{(1-R_{\mathrm{p}}-\gamma)(1-\epsilon_0)}$, which is nonnegative by assumption, into equation (62) to get that

$$\left( \frac{(R_{\mathrm{p}}+\gamma)\epsilon_0}{(1-R_{\mathrm{p}}-\gamma)(1-\epsilon_0)} \right)^s \geq \left( \frac{(R_{\mathrm{p}}+\gamma)\epsilon_0}{(1-R_{\mathrm{p}}-\gamma)(1-\epsilon_0)} \right)^{2r} \mathbb{E}[e^{(N \wedge m)D(1-R_{\mathrm{p}}-\gamma\|\epsilon_0)}|S_0 = s] , \quad (63)$$

where the inequality follows since $S_k$ has bounded increments that are less than 1 almost surely and the stopping time that increases in multiples of $r$ increments. By the monotone convergence [21, p. 15, Theorem 1.3.6], letting $m \to \infty$ gives

$$\mathbb{E}[e^{ND(1-R_{\mathrm{p}}-\gamma\|\epsilon_0)}|S_0 = s] \leq \left( \frac{(1-R_{\mathrm{p}}-\gamma)(1-\epsilon_0)}{(R_{\mathrm{p}}+\gamma)\epsilon_0} \right)^{2r-s} . \quad (64)$$

Using this moment inequality, we can apply a Chernoff bound to conclude

$$\mathbb{P}(N \geq t|S_0 = s) \leq \left( \frac{(1-R_{\mathrm{p}}-\gamma)(1-\epsilon_0)}{(R_{\mathrm{p}}+\gamma)\epsilon_0} \right)^{2r-s} e^{-tD(1-R_{\mathrm{p}}-\gamma\|\epsilon_0)} . \quad (65)$$

∎

*Lemma 3:* Given a strategy in which for all $n$, every code of blocklength $n$ satisfies (18), if $\gamma > 0$ is chosen so that $\gamma < \frac{\nu}{2}$, then the strategy is valid. That is, for all $\{\epsilon_{k,1}\}_{i=1}^{\infty}$,

$$\mathbb{P}_{\epsilon_0, \{\epsilon_{k,1}\}_{i=1}^{\infty}} \left( \ell^{-1} \sum_{k=1}^{\ell} A_k < R_{\mathrm{p}} \right) \leq K_{1,R_{\mathrm{p}},\nu,\ell} e^{-\ell K_{2,R_{\mathrm{p}},\nu}} , \quad (66)$$

where $0 < K_{2,R_{\mathrm{p}},\nu} < \infty$, $K_{1,R_{\mathrm{p}},\nu,\ell} < \infty$, and $K_{1,R_{\mathrm{p}},\nu,\ell} e^{-\ell K_{2,R_{\mathrm{p}},\nu}} \to 0$ as $\ell \to \infty$.

*Proof:* When $S_\ell > 0$, the primary is meeting its rate target. Furthermore, if $S_{iK_n} - K_n - iK_n\gamma > 0$, then the primary will be guaranteed to meet its rate target over the next primary frame. Thus, it suffices to consider frames when $S_{iK_n} - K_n - iK_n\gamma \leq 0$, which correspond directly to silent frames. To consider what happens in these settings, we define stopping times to threshold $S_\ell - \ell\gamma$.

$$N_{2k-1} = \inf\{\ell > N_{2k-2} : S_\ell - \ell\gamma \geq 1\}, \quad (67)$$

$$N_{2k} = \inf\{\ell > N_{2k-1} : S_\ell - \ell\gamma < 1\}. \quad (68)$$

Negative deviations occur only when $N_{2k} \leq \ell < N_{2k+1}$, so

$$\mathbb{P}(S_\ell \leq 0)$$

$$\leq \sum_k \mathbb{P}(-S_\ell + \ell\gamma \geq \ell\gamma | N_{2k} \leq \ell < N_{2k+1})\mathbb{P}(N_{2k} \leq \ell < N_{2k+1}) \quad (69)$$

$$\leq \sum_k \mathbb{P}(N_{2k+1} - N_{2k} \geq \ell\gamma | N_{2k} \leq \ell < N_{2k+1})\mathbb{P}(N_{2k} \leq \ell < N_{2k+1}) \quad (70)$$

$$\leq \sum_k \mathbb{P}(N_{2k+1} - N_{2k} \geq \ell\gamma | N_{2k} < \infty)\mathbb{P}(N_{2k} \leq \ell < N_{2k+1} | N_{2k+1} - N_{2k} \geq \ell\gamma, N_{2k} < \infty) \quad (71)$$

$$\leq \frac{\ell}{2} \max_{1 \leq 2k \leq \ell} \mathbb{P}(N_{2k+1} - N_{2k} \geq \ell\gamma | N_{2k} < \infty) , \quad (72)$$

where (69) follows from the law of total probability; (70) since the bounded increments of $S_\ell$ imply that $-S_\ell + \ell\gamma \leq N_{2k+1} - N_{2k}$ given $N_{2k} \leq \ell < N_{2k+1}$; (71) from Bayes theorem and that probabilities are bounded from above by 1; and (72) since probabilities are bounded from above by 1 and for $N_{2k} \leq \ell$, it is necessary for $2k \leq \ell$ by the definition of $N_{2k}, N_{2k+1}$. Since $S_n$ is a Markov chain, then for all $k$, Lemma 13 implies that we only need to consider

$$\mathbb{P}(N_{2k+1} - N_{2k} \geq \ell\gamma | N_{2k} < \infty) \leq \max_{s \in [-1,0]} \mathbb{P}(N_1 \geq \ell\gamma | S_0 = s) . \tag{73}$$

$$\leq \left( \frac{(1 - R_\mathrm{p} - \gamma)(1 - \epsilon_0)}{(R_\mathrm{p} + \gamma)\epsilon_0} \right)^2 e^{-\ell D(1 - R_\mathrm{p} - \gamma \| \epsilon_0)} \tag{74}$$

where (74) follows from Lemma 2. Since the above holds for all $k$, substituting it into (72) gives

$$\mathbb{P}(S_\ell \leq 0)$$

$$\leq \frac{\ell}{2} \left( \frac{(1 - R_\mathrm{p} - \gamma)(1 - \epsilon_0)}{(R_\mathrm{p} + \gamma)\epsilon_0} \right)^2 e^{-\ell D(1 - R_\mathrm{p} - \gamma \| \epsilon_0)} \tag{75}$$

$$= \frac{\ell}{2} \cdot \exp\left( \frac{2}{1 - R_\mathrm{p} - \gamma} \cdot \left( D(1 - R_\mathrm{p} - \gamma \| \epsilon_0) - \log \frac{R_\mathrm{p} + \gamma}{1 - \epsilon_0} \right) - \ell D(1 - R_\mathrm{p} - \gamma \| \epsilon_0) \right) \tag{76}$$

$$= \frac{\ell}{2} \cdot \exp\left( -\frac{2}{1 - R_\mathrm{p} - \gamma} \cdot \log \frac{R_\mathrm{p} + \gamma}{1 - \epsilon_0} - (\ell - 2/(1 - R_\mathrm{p} - \gamma)) D(1 - R_\mathrm{p} - \gamma \| \epsilon_0) \right) \tag{77}$$

$$\leq \frac{\ell}{2} \cdot \exp\left( -\frac{2}{1 - R_\mathrm{p} - \gamma} \cdot \log \frac{R_\mathrm{p} + \gamma}{R_\mathrm{p} + \nu} - (\ell - 2/(1 - R_\mathrm{p} - \gamma)) D(1 - R_\mathrm{p} - \gamma \| \epsilon_0) \right) , \tag{78}$$

where the last line follows since $1 - \epsilon_0 \geq R_\mathrm{p} + \nu$. For $\ell > 2/(1 - R_\mathrm{p} - \gamma)$, we use the fact that $D(1 - R_\mathrm{p} - \gamma \| \epsilon_0) \geq \frac{(1 - R_\mathrm{p} - \gamma - \epsilon_0)^2}{2}$ to get that

$$\mathbb{P}(S_\ell \leq 0)$$

$$\leq \frac{\ell}{2} \cdot \exp\left( -\frac{2}{1 - R_\mathrm{p} - \gamma} \cdot \log \frac{R_\mathrm{p} + \gamma}{R_\mathrm{p} + \nu} - (\ell - 2/(1 - R_\mathrm{p} - \gamma)) \frac{(1 - R_\mathrm{p} - \gamma - \epsilon_0)^2}{2} \right) \tag{79}$$

$$\leq \frac{\ell}{2} \cdot \exp\left( -\frac{2}{1 - R_\mathrm{p} - \gamma} \cdot \log \frac{R_\mathrm{p} + \gamma}{R_\mathrm{p} + \nu} - (\ell - 2/(1 - R_\mathrm{p} - \gamma)) \frac{\nu^2}{8} \right) , \tag{80}$$

where the last line follows by assumption. Note that this expression goes to 0 as $\ell \to \infty$. By letting $K_{2,R_\mathrm{p},\nu} = \frac{\nu^2}{8}$, and $K_{1,R_\mathrm{p},\nu,\ell}$ to be (80) divided by $e^{-\ell K_{2,R_\mathrm{p},\nu}}$ for $\ell > 2/(1 - R_\mathrm{p} - \gamma)$, then we can write

$$\mathbb{P}(S_\ell \leq 0) \leq K_{1,R_\mathrm{p},\nu,\ell} e^{-\ell K_{2,R_\mathrm{p},\nu}} . \tag{81}$$

For $\ell \leq 2/(1 - R_\mathrm{p} - \gamma)$, we simply choose $K_{1,R_\mathrm{p},\nu,\ell}$ to make the probability upper bound 1. Then we can conclude our result by simply recalling the definition of $S_\ell$ in (17). ■

### B. Decoder Error

*Lemma 4:* Define error events as follows:

1) $E_1$: $\exists$ frame in which the decoder misidentifies it as active or silent.

2) $E_2$: $\exists$ frame in which the decoder misidentifies the codeword.

Then, for $\tilde{\delta} > 0, \kappa_n = \lfloor n^{1/16} \rfloor, K_n = \lfloor n^{1/8} \rfloor$ in the fixed-codebook protocol, as $n \to \infty$,

$$\mathbb{P}(E_1 \cup E_2) \to 0 \ . \tag{82}$$

*Proof:* Note that we can bound the error as

$$\mathbb{P}(E_1 \cup E_2) \le \mathbb{P}(E_1) + \mathbb{P}(E_2|E_1^c) \ . \tag{83}$$

First consider the event of misidentifying whether transmission is taking place over a frame. There are $\frac{n}{K_n}$ frames, and an error occurs if misidenification happens over any one of them. Thus, by taking a union bound over all frames and applying Lemma 15 for the error of the repetition code used to distinguish these frames, the error probability is bounded by

$$\mathbb{P}(E_1) \le \frac{n}{K_n} e^{-\kappa_n \cdot r} \ , \tag{84}$$

where $r > 0$ is independent of $n$.

Finally, we can consider the error corresponding to misidentifying the codewords sent in each frame. By Lemma 14 and a union bound per frame, we also also have that

$$\mathbb{P}(E_2|E_1^c) \le \frac{n}{K_n} \cdot e^{-\frac{(K_n - \kappa_n)\delta^2}{8/\epsilon^2 + 4(\log |\mathcal{Y}|)^2}} \ . \tag{85}$$

Combining (84) and (85) with (83), we complete the proof. ∎

### C. Rate Analysis

The rate achievable by the cognitive radio is directly proportional to the fraction of frames in which it is active. Thus, we consider a bound on the number of frames the primary is guaranteed to be active.

*Lemma 5:* For all $\delta > 0$ and $\gamma < \delta/2$, there exists an $n_0(\delta)$ such that for $n \ge n_0(\delta)$,

$$\mathbb{P}\left(n^{-1}\sum_{k=1}^{n}\tau_k \le \frac{1 - R_{\mathsf{p}} - \epsilon_0}{1 - \epsilon_0} - \delta\right) \le \exp\left\{-nD\left(\frac{1 + \delta/2}{2} - n^{-1}K_n \middle\| \frac{1}{2}\right)\right\} \ , \tag{86}$$

where $\tau_k$ is an indicator random variable to denote that $k$ is in an active frame. That is,

$$\tau_k = \mathbb{I}_{\{\exists \ell \in \mathbb{Z}^+ \text{ such that } V_\ell < k \le V_\ell + K_n\}} \ ,$$

where $V_\ell$ is defined in (21). Furthermore, if $K_n = o(n)$, (86) goes to 0 as $n \to \infty$.

*Proof:* Note that if the sequence $\epsilon_{x,k} = 1$ for all $k$, $x \neq x_{\text{off}}$, and the transmitter does not use the symbol $x_{\text{off}}$ during active frames, all channel uses in an active frame result in an erased packet for the primary. Thus, given $\epsilon_0$, this case provides a convenient albeit conservative way to lower bound the fraction of active frames. We will assume it in the sequel.

Recalling the definition of $S_n$ in (17) and the definition of $\tau_k$ in (18), $S_n$ can be no more than $n\gamma + 2K_n$ during the course of a silent frame. Furthermore, it can only decrease from this during an active frame since $\epsilon_{x,k} = 1$, $x \neq x_{\text{off}}$. Thus, we are guaranteed almost surely that

$$S_n \leq n\gamma + 2K_n \ . \tag{87}$$

Recalling the definition of $S_n$ in (17), we can define

$$\tilde{S}_n = S_n + \sum_{k=1}^{n} \tau_k (1 - \epsilon_0) - n(1 - R_{\mathfrak{p}} - \epsilon_0) \tag{88}$$

Thus, we can rewrite the problem as showing that

$$\mathbb{P}\left( \sum_{k=1}^{n} \tau_k (1 - \epsilon_0) - n(1 - R_{\mathfrak{p}} - \epsilon_0) \leq -n\delta \right)$$

$$= \mathbb{P}(\tilde{S}_n - S_n \leq -n\delta) \tag{89}$$

$$= \mathbb{P}(\tilde{S}_n \leq S_n - n\delta) \tag{90}$$

$$\leq \mathbb{P}(\tilde{S}_n \leq -n\delta/2 + 2K_n) \tag{91}$$

is small, where the last line follows from (87) and the assumption that $\gamma < \delta/2$.

It is straightforward to verify that for $\epsilon_{x,k} = 1$, $x \neq x_{\text{off}}$, $\tilde{S}_n$ is a martingale by checking that $\mathbb{E}[\tilde{S}_n | \tilde{S}_0, \dots, \tilde{S}_{n-1}] = \tilde{S}_{n-1}$. Furthermore, it has bounded increments. Thus, a bounded martingale concentration inequality [22, p. 57, Corollary 2.4.7] implies that for large enough $n$,

$$\mathbb{P}(\tilde{S}_n \leq -n\delta/2 + 2K_n) \leq \exp\left\{ -nD\left( \frac{1 + \delta/2}{2} - n^{-1}K_n \bigg\| \frac{1}{2} \right) \right\} \ . \tag{92}$$

The result follows immediately. ∎

*Lemma 6:* Given $\delta > 0$, $\epsilon_0$, $\{\epsilon_{1,k}\}_{k=1}^{\infty}$, consider the fixed-rate protocol with $\kappa_n = o(K_n), K_n = o(n), K_n \to \infty$ as $n \to \infty$, and $\gamma < \tilde{\delta}/2$. Then there exists a choice of $\tilde{\delta}$ such that this strategy attains rates of at least

$$\left( 1 - \frac{R_{\mathfrak{p}}}{1 - \epsilon_0} \right) \cdot C^* - \delta \ , \tag{93}$$

with probability going to 1 as $n \to \infty$, where $C^* = \max_{p(x)} I(X; Y)$.

*Proof:* By Lemma 5, we know that with probability going to 1 as $n \to \infty$, at least

$$\left( \frac{1 - R_\mathrm{p} - \epsilon_0}{1 - \epsilon_0} - \tilde{\delta} \right) n \tag{94}$$

of the frames will be active frames. The rate for each active frame is

$$\frac{K_n - \kappa_n}{K_n} (C^* - \tilde{\delta}) , \tag{95}$$

so we have that

$$\left( \frac{1 - R_\mathrm{p} - \epsilon_0}{1 - \epsilon_0} - \tilde{\delta} \right) \cdot \frac{K_n - \kappa_n}{K_n} (C^* - \tilde{\delta})$$
$$\geq \left( \frac{1 - R_\mathrm{p} - \epsilon_0}{1 - \epsilon_0} \right) C^* - \tilde{\delta} - (\kappa_n / K_n + \tilde{\delta}) \cdot C^* . \tag{96}$$

For large enough n, by assumption $\kappa_n / K_n \leq \tilde{\delta} / C^*$, so that rates of at least

$$\left( \frac{1 - R_\mathrm{p} - \epsilon_0}{1 - \epsilon_0} \right) C^* - 2\tilde{\delta} - \tilde{\delta} \cdot C^* \tag{97}$$

are achievable with probability going to 1. By choosing $\tilde{\delta} = \frac{1}{3} \min\{\delta, \delta / C^*\}$, we can conclude the result.
∎

## APPENDIX II

## PROOF OF THEOREM 2

### A. Decoder Error

*Lemma 7:* Define error events as follows:

1) $E_1$: ∃ frame in which the decoder misidentifies whether a frame is active or silent.

2) $E_2$: the decoder misidentifies the selected codebook.

3) $E_3$: ∃ an active frame in which decoder misidentifies the codeword.

Then, for $\tilde{\delta} > 0, C_n = \lfloor n^{1/32} \rfloor, \kappa_n = \lfloor n^{1/16} \rfloor, K_n = \lfloor n^{1/8} \rfloor$ in the codebook-adaptive protocol, as $n \to \infty$,

$$\mathbb{P}(E_1 \cup E_2 \cup E_3) \to 0 . \tag{98}$$

*Proof:* Note that we can bound the error as

$$\mathbb{P}(E_1 \cup E_2 \cup E_3) \leq \mathbb{P}(E_1) + \mathbb{P}(E_2 | E_1^c) + \mathbb{P}(E_3 | E_1^c, E_2^c) . \tag{99}$$

First consider the event of misidentifying whether transmission is taking place over a frame. There are $\frac{n}{K_n}$ frames, and an error occurs if misidentification happens over any one of them. Thus, by taking a union bound over all frames and applying Lemma 15 for the error of the repetition code used to distinguish these frames, the error probability is bounded by

$$\mathbb{P}(E_1) \leq \frac{n}{K_n} e^{-\kappa_n \cdot r} , \tag{100}$$

where $r > 0$ is independent of $n$. Another source of error is misidentifying the codebook. For large enough $n$, $(C_n + 1)^{|\mathcal{X}|}$ does not exceed $C - \delta$, and we can apply Lemma 14 to get the error probability

$$\mathbb{P}(E_2 | E_1^c) \leq e^{-\frac{(K_n - \kappa_n) \cdot \hat{\delta}^2}{8/\epsilon^2 + 4(\log |\mathcal{Y}|)^2}} . \tag{101}$$

Finally, we can consider the error corresponding to misidentifying the codewords sent in each frame. By Lemma 14 and a union bound per frame, we also also have that

$$\mathbb{P}(E_3 | E_2^c, E_1^c) \leq \frac{n}{K_n} \cdot e^{-\frac{(K_n - \kappa_n) \hat{\delta}^2}{8/\epsilon^2 + 4(\log |\mathcal{Y}|)^2}} . \tag{102}$$

Combining (100), (101), and (102) with (99), we complete the proof. ∎

## B. Rate Analysis

The rate loss argument is the most tedious because one must account for a variety of factors: the length of the first two phases of transmission, the gap between the rates of quantized set of codebooks and points on the RIB function, and the number of active frames in Phase III. We therefore subdivide the result into several lemmas.

*1) Phase I and II are short:* Because the encoder does not send message information in Phase I and II, we want the length of these phases to be sublinear in $n$ to guarantee negligible rate loss.

*Lemma 8:* For all $\nu > 0$, let $\gamma < \nu/2$, $\kappa_n = \lfloor n^{1/16} \rfloor$, $K_n = \lfloor n^{1/8} \rfloor$ in the codebook-adaptive protocol. Furthermore, let $T$ be the length of Phases I and II and $\tilde{E}_1 = \{T \geq n^{1/4}\}$. Then

$$\mathbb{P}(\tilde{E}_1) \to 0 \tag{103}$$

as $n \to \infty$.

*Proof:* Consider the transition times from silent frames to active frames and vice versa. To do this, define the stopping times for $k \geq 1$,

$$\tilde{N}_{2k-1} = K_n \cdot \inf\{i > K_n^{-1} \tilde{N}_{2k-2} : S_{i \cdot K_n} - i \cdot K_n \cdot \gamma - K_n \geq 0\} , \tag{104}$$

$$\tilde{N}_{2k} = K_n \cdot \inf\{i > K_n^{-1} \tilde{N}_{2k-1} : S_{i \cdot K_n} - i \cdot K_n \cdot \gamma - K_n < 0\} , \tag{105}$$

where $\tilde{N}_0 = 0$. Phases I and II end after the first two active frames. We can get a bound on the start of the first active frame immediately from Lemma 2, which implies

$$\mathbb{P}(\tilde{N}_1 \geq t | S_0 = 0) \leq \left( \frac{(1 - R_{\mathsf{p}} - \gamma)(1 - \epsilon_0)}{(R_{\mathsf{p}} + \gamma)\epsilon_0} \right)^{2K_n} e^{-tD(1 - R_{\mathsf{p}} - \gamma \| \epsilon_0)} . \quad (106)$$

Thus, if $\tilde{N}_2 > \tilde{N}_1 + K_n$, then

$$T = \tilde{N}_1 + 2K_n. \quad (107)$$

Together with (106), this implies

$$\mathbb{P}(T \geq n^{1/4} | \tilde{N}_2 > \tilde{N}_1 + K_n) \leq \left( \frac{(1 - R_{\mathsf{p}} - \gamma)(1 - \epsilon_0)}{(R_{\mathsf{p}} + \gamma)\epsilon_0} \right)^{2K_n} e^{-(n^{1/4} - K_n) \cdot D(1 - R_{\mathsf{p}} - \gamma \| \epsilon_0)} \quad (108)$$

The remaining case to consider is if $\tilde{N}_2 = \tilde{N}_1 + K_n$. If this happens, then Phases I and II end at

$$T = \tilde{N}_3 + K_n. \quad (109)$$

Then (109) implies

$$\mathbb{P}(T \geq n^{1/4} | \tilde{N}_2 = \tilde{N}_1 + K_n) = \mathbb{P}(\tilde{N}_3 \geq n^{1/4} - K_n | \tilde{N}_2 = \tilde{N}_1 + K_n) \quad (110)$$

$$= \mathbb{P}(\tilde{N}_3 - \tilde{N}_2 \geq n^{1/4} - 2K_n - \tilde{N}_1 | \tilde{N}_2 = \tilde{N}_1 + K_n) \quad (111)$$

$$\leq \mathbb{P}(\tilde{N}_3 - \tilde{N}_2 \geq n^{1/4} - 2K_n - \tilde{N}_1 \text{ or } \tilde{N}_1 \geq n^{1/4}/2 | \tilde{N}_2 = \tilde{N}_1 + K_n) \quad (112)$$

$$\leq \mathbb{P}(\tilde{N}_1 \geq n^{1/4}/2 | \tilde{N}_2 = \tilde{N}_1 + K_n)$$
$$+ \mathbb{P}(\tilde{N}_3 - \tilde{N}_2 \geq n^{1/4}/2 - 2K_n | \tilde{N}_2 = \tilde{N}_1 + K_n, \tilde{N}_1 < n^{1/4}/2) , \quad (113)$$

where (110) follows from (109), (111) follows by our conditioning, (112) follows since we are increasing the possible events over which we are taking the probability, and (113) follows from $P(A \text{ or } B) = P(A) + P(A^c)P(B|A^c)$.

By Lemma 13 and Lemma 2,

$$\mathbb{P}(\tilde{N}_3 - \tilde{N}_2 \geq n^{1/4}/2 - 2K_n | \tilde{N}_2 = \tilde{N}_1 + K_n, \tilde{N}_1 < n^{1/4}/2)$$
$$\leq \left( \frac{(1 - R_{\mathsf{p}} - \gamma)(1 - \epsilon_0)}{(R_{\mathsf{p}} + \gamma)\epsilon_0} \right)^{2K_n} e^{-(n^{1/4}/2 - 2K_n) \cdot D(1 - R_{\mathsf{p}} - \gamma \| \epsilon_0)} . \quad (114)$$

By combining (106), and (113), and (114),

$$\mathbb{P}(T \geq n^{1/4} | \tilde{N}_2 = \tilde{N}_1 + K_n)$$

$$\leq \left( \frac{(1 - R_\mathsf{p} - \gamma)(1 - \epsilon_0)}{(R_\mathsf{p} + \gamma)\epsilon_0} \right)^{2K_n} e^{-(n^{1/4}/2 - 2K_n) \cdot D(1 - R_\mathsf{p} - \gamma \| \epsilon_0)}$$

$$+ \left( \frac{(1 - R_\mathsf{p} - \gamma)(1 - \epsilon_0)}{(R_\mathsf{p} + \gamma)\epsilon_0} \right)^{2K_n} e^{-\frac{n^{1/4}}{2} D(1 - R_\mathsf{p} - \gamma \| \epsilon_0)} . \tag{115}$$

The result follows immediately from (108) and (115). ∎

*2) Codebook Quantization:* To account for the error in the interference estimates and that the encoder must inform the decoder of which rate it will be targetting, we only have a limited number of codebooks to choose from at the start of Phase II. Thus, in general there will be a gap between the rate of a selected codebook and an actual point on the RIB function. In this subsection, we ensure that this gap is small.

We first provide guarantees on accurate interference estimates.

*Lemma 9:* Let $\mathcal{D}_\ell$ be the event that Phase I of the codebook-adaptive protocol terminates at frame $\ell$. Then

$$\mathbb{P}(\max_x |\hat{\epsilon}_x - \epsilon_x| > \delta | \mathcal{D}_\ell) \leq 2|\mathcal{X}| e^{-\mu \delta^2 / 2} , \tag{116}$$

where $\mu = \lfloor \frac{K_n - \kappa_n}{|\mathcal{X}|} \rfloor$.

*Proof:* Recall the definition of the estimates given in (29). Then $\mathcal{D}_\ell = \{V_1 = (\ell - 1)K_n\}$ is an equivalent expression for the event. By Hoeffding's inequality [22, p. 57, Corollary 2.4.7], we have for each $x \in \mathcal{X}$

$$\mathbb{P}(|\hat{\epsilon}_x - \epsilon_x| > \delta | V_1 = (\ell - 1)K_n) \leq 2e^{-\mu \delta^2 / 2} . \tag{117}$$

The result then follows from a union bound on $\mathbb{P}(\max_x |\hat{\epsilon}_x - \epsilon_x| > \delta | \mathcal{D}_\ell)$. ∎

*Lemma 10:* For the selected codebook $\chi$ given in (31), define $\tilde{E}_2^c$ as the event where the following two conditions are met:

$$\sum_x \epsilon_x p_\chi(x) \leq 1 - R_\mathsf{p} - 2\gamma \tag{118}$$

$$6\frac{|\mathcal{X}|}{C_n} \log \frac{2}{C_n \cdot |\mathcal{Y}|} + \frac{3\tilde{\delta}}{2|\mathcal{X}|} \log \frac{\tilde{\delta}}{2|\mathcal{X}|^2 \cdot |\mathcal{Y}|}$$
$$\leq R_\chi - R_{\mathrm{IB}}(\vec{\epsilon}, R_\mathsf{p} + 2\gamma + \tilde{\delta}) \leq$$
$$- \frac{3\tilde{\delta}}{2|\mathcal{X}|} \log \frac{\tilde{\delta}}{2|\mathcal{X}|^2 \cdot |\mathcal{Y}|}. \tag{119}$$

 

Then for all $1 > \tilde{\delta} > 0$, $C_n \geq 4|\mathcal{X}|$, and as $n \to \infty$,

$$\mathbb{P}(\tilde{E}_2|\tilde{E}_1^c) \to 0 \ , \tag{120}$$

where $\tilde{E}_1$ is defined in Lemma 8.

*Proof:* Given $\tilde{E}_1^c$, we can guarantee by setting $\delta = \frac{\tilde{\delta}}{4|\mathcal{X}|}$ in Lemma 9 that

$$\sum_x |\hat{\epsilon}_x - \epsilon_x| \leq \frac{\tilde{\delta}}{4} \tag{121}$$

with probability going to 1 as $n \to \infty$. Furthermore, we know that by definition

$$\sum_x \hat{\epsilon}_x p_\chi(x) \leq 1 - R_{\mathrm{p}} - 2\gamma - \tilde{\delta} \tag{122}$$

From (121) and (122), we have that

$$\sum_x \epsilon_x p_\chi(x) \leq \sum_x \hat{\epsilon}_x p_\chi(x) + \tilde{\delta} \tag{123}$$

$$\leq 1 - R_{\mathrm{p}} - 2\gamma. \tag{124}$$

It remains to verify the other condition. Note that for any $p(x)$, there is a codebook in the set with input distribution type $p_{x^{C_n}}(x)$ such that $\sum_x |p(x) - p_{x^{C_n}}(x)| \leq \frac{|\mathcal{X}|}{C_n}$. Then by the continuity of entropy [23, Lemma 2.7, p. 33], we know that

$$R_{\mathrm{IB}}(\vec{\tilde{\epsilon}}, R_{\mathrm{p}} + 2\gamma + \tilde{\delta}) + 6\frac{|\mathcal{X}|}{C_n} \log \frac{2}{C_n \cdot |\mathcal{Y}|} \leq R_\chi \leq R_{\mathrm{IB}}(\vec{\tilde{\epsilon}}, R_{\mathrm{p}} + 2\gamma + \tilde{\delta}) \ , \tag{125}$$

where the inequality on the right follows from (122) and the definition of the $R_{\mathrm{IB}}$ function. Lemma 18 and (121) imply that

$$|R_{\mathrm{IB}}(\vec{\epsilon}, R_{\mathrm{p}} + 2\gamma + \tilde{\delta}) - R_{\mathrm{IB}}(\vec{\tilde{\epsilon}}, R_{\mathrm{p}} + 2\gamma + \tilde{\delta})| \leq -\frac{3\tilde{\delta}}{2|\mathcal{X}|} \log \frac{\tilde{\delta}}{2|\mathcal{X}|^2 \cdot |\mathcal{Y}|} \ . \tag{126}$$

Combining (125) and (126) yield the result. ∎

*3) Always On:* Our next lemma shows that all frames are active after time $\sqrt{n}$ with probability going to 1 as $n \to \infty$.

*Lemma 11:* Let $\tilde{E}_1$ and $\tilde{E}_2$ be defined as in Lemmas 8 and 10, respectively. Define $\tilde{E}_3$ to be the event that for some $j > \sqrt{n}$, the condition in (26) is met, resulting in a silent frame. Then for all $\nu > 0$ and $\epsilon_{x,i} = \epsilon_x$ for $x \neq x_{\mathrm{off}}$, the codebook-adaptive protocol with parameters $(\gamma, C_n, \kappa_n, K_n, \tilde{\delta})$ satisfying $0 < \gamma < \nu/2, C_n = \lfloor n^{1/32} \rfloor, \kappa_n = \lfloor n^{1/16} \rfloor, K_n = \lfloor n^{1/8} \rfloor, 1 > \tilde{\delta} > 0$, has the property that as $n \to \infty$,

$$\mathbb{P}(\tilde{E}_3|\tilde{E}_1^c, \tilde{E}_2^c) \to 0 \ . \tag{127}$$

*Proof:* We start by defining

$$\tilde{S}_j = \tilde{S}_{j-1} + A_j - \mathbb{E}[A_j|\tilde{S}_0, \ldots, \tilde{S}_{j-1}], \quad \tilde{S}_0 = 0, \tag{128}$$

and it is easy to verify that $\tilde{S}_j$ is a bounded martingale. From the definition of $\tilde{E}_2^c$ in Lemma 10 and for $\nu > 0$, (118), (26), and (33) imply that for $j$ satisfying $j > V_2 + K_n$ and $\ell K_n \geq j > (\ell-1)K_n + \kappa_n$ for some integer $\ell \geq 1$ ,

$$\mathbb{E}[A_j|\tilde{S}_0, \ldots, \tilde{S}_{j-1}] \geq R_{\mathfrak{p}} + 2\gamma \ . \tag{129}$$

Then for $k \geq n^{1/2}$ and under $\tilde{E}_1^c$,

$$\mathbb{P}\left( k^{-1} \sum_{i=1}^k A_i < R_{\mathfrak{p}} + \gamma + k^{-1}K_n \,\middle|\, \tilde{E}_1^c, \tilde{E}_2^c \right)$$

$$\leq \mathbb{P}\left( k^{-1}\tilde{S}_k < R_{\mathfrak{p}} + \gamma + k^{-1}K_n - k^{-1}\beta_{k,n}(R_{\mathfrak{p}} + 2\gamma) \,\middle|\, \tilde{E}_1^c, \tilde{E}_2^c \right) \tag{130}$$

$$\leq \mathbb{P}\left( k^{-1}\tilde{S}_k < -\gamma + o(1) \,\middle|\, \tilde{E}_1^c, \tilde{E}_2^c \right) \ , \tag{131}$$

where $\beta_{k,n} = (k - n^{1/4}) \cdot \frac{K_n - \kappa_n}{K_n}$, $o(1)$ is notational convenience for $\lim_{n \to \infty} o(1) = 0$, and (130) follows from (128), (129), and since for all $i$, $A_i \geq 0$ almost surely. Since $\tilde{S}_k$ is a zero-mean bounded martingale, for $k \geq n^{1/2}$ and large enough $n$, we can apply a bounded martingale concentration inequality [22, p. 57, Corollary 2.4.7] to yield

$$\mathbb{P}\left( k^{-1} \sum_{i=1}^k A_k < R_{\mathfrak{p}} + \gamma + K_n \,\middle|\, \tilde{E}_1^c, \tilde{E}_2^c \right)$$

$$\leq \exp\left( -k(\gamma + o(1))^2/2 \right) \tag{132}$$

$$= \exp\left( -\lceil\sqrt{n}\rceil(\gamma + o(1))^2/2 \right) \cdot \exp\left( -(k - \lceil\sqrt{n}\rceil)(\gamma + o(1))^2/2 \right) \tag{133}$$

From the above result and a union bound,

$$\mathbb{P}(\tilde{E}_3|\tilde{E}_1^c, \tilde{E}_2^c)$$

$$\leq \exp\left( -\lceil\sqrt{n}\rceil(\gamma + o(1))^2/2 \right) \cdot \sum_{k=\lceil\sqrt{n}\rceil}^n \exp\left( -(k - \lceil\sqrt{n}\rceil)(\gamma + o(1))^2/2 \right) \tag{134}$$

$$\leq \exp\left( -\lceil\sqrt{n}\rceil(\gamma + o(1))^2/2 \right) \cdot \sum_{m=0}^\infty \exp\left( -m(\gamma + o(1))^2/2 \right) \tag{135}$$

However, the geometric series does not affect the error probability by more than a constant asymptotically, so taking the limit above completes the result. ∎

*4) Overall Rate Loss:*

*Lemma 12:* For all $\nu > 0$, $\epsilon_{x,i} = \epsilon_x$ for $x \neq x_{\text{off}}$ and given any $\delta > 0$, consider the codebook-adaptive protocol with parameters $(\gamma, C_n, \kappa_n, K_n)$ satisfying $0 < \gamma < \min\{\nu/2, \tilde{\delta}/2\}$, $C_n = \lfloor n^{1/32} \rfloor$, $\kappa_n = \lfloor n^{1/16} \rfloor$, $K_n = \lfloor n^{1/8} \rfloor$. Then there exists a choice of the parameter $\tilde{\delta} \in (0, 1/8)$ so that with probability going to 1 as $n \to \infty$, the cognitive radio achieves rates

$$R \geq R_{\text{IB}}(\vec{\epsilon}, R_{\text{p}}) - \delta \ . \tag{136}$$

*Proof:* Let $\tilde{E}_1, \tilde{E}_2, \tilde{E}_3$ be defined as in Lemmas 8, 10, and 11 respectively. From these results, we know that

$$\mathbb{P}(\tilde{E}_1 \cup \tilde{E}_2 \cup \tilde{E}_3) \to 0 \tag{137}$$

as $n \to \infty$ and thus with high probability,

1) Phases I and II are short, ending by $n^{1/4}$ (Lemma 8).

2) The gap between the codebook's rate and the RIB function is small (Lemma 10).

3) After time $n^{1/2}$, all frames are active frames (Lemma 11).

Furthermore, we know that by our repetition code, there is a loss of $\kappa_n$ positions for our repetition code over a frame $K_n$. Factoring in this source of rate loss along with the fact that we are in Phase III by time $n^{1/2}$ (Lemma 8) and always in an active frame (Lemma 11), the rate

$$\frac{n - \sqrt{n}}{n} \cdot \frac{K_n - \kappa_n}{K_n}(R_\chi - \tilde{\delta})$$
$$\geq (R_\chi - \tilde{\delta}) - \left( n^{-1/2} + \frac{\kappa_n}{K_n} \right) \log |\mathcal{X}| \tag{138}$$

is achievable for the cognitive radio with probability going to 1 as $n \to \infty$. Finally, we know that the gap between the codebook's rate and the RIB function is small (Lemma 10), so

$$R_\chi \geq R_{\text{IB}}(\vec{\epsilon}, R_{\text{p}} + 2\gamma + \tilde{\delta}) + 6\frac{|\mathcal{X}|}{C_n} \log \frac{2}{C_n \cdot |\mathcal{Y}|} + \frac{3\tilde{\delta}}{2|\mathcal{X}|} \log \frac{\tilde{\delta}}{2|\mathcal{X}|^2 \cdot |\mathcal{Y}|} \tag{139}$$

$$\geq R_{\text{IB}}(\vec{\epsilon}, R_{\text{p}} + 2\tilde{\delta}) + 6\frac{|\mathcal{X}|}{C_n} \log \frac{2}{C_n \cdot |\mathcal{Y}|} + \frac{3\tilde{\delta}}{2|\mathcal{X}|} \log \frac{\tilde{\delta}}{2|\mathcal{X}|^2 \cdot |\mathcal{Y}|} \ , \tag{140}$$

$$\geq R_{\text{IB}}(\vec{\epsilon}, R_{\text{p}}) + 6\frac{|\mathcal{X}|}{C_n} \log \frac{2}{C_n \cdot |\mathcal{Y}|} + \frac{3\tilde{\delta}}{2|\mathcal{X}|} \log \frac{\tilde{\delta}}{2|\mathcal{X}|^2 \cdot |\mathcal{Y}|} + 12\tilde{\delta} \log \frac{4\tilde{\delta}}{|\mathcal{X}| \cdot |\mathcal{Y}|} \ , \tag{141}$$

where (140) follows by our assumption about $\gamma$ and (141) from Lemma 17 given our assumption about $\tilde{\delta}$. Combining (138) with (141), our assumptions about $(C_n, \kappa_n, K_n)$ imply that for large enough $n$, the rate

$$R_{\text{IB}}(\vec{\epsilon}, R_{\text{p}}) - \frac{\delta}{2} - \left( \tilde{\delta} + \frac{3\tilde{\delta}}{2|\mathcal{X}|} \log \frac{2|\mathcal{X}|^2 \cdot |\mathcal{Y}|}{\tilde{\delta}} + 12\tilde{\delta} \log \frac{|\mathcal{X}| \cdot |\mathcal{Y}|}{4\tilde{\delta}} \right) \tag{142}$$

is achievable for the cognitive radio with probability going to 1 as $n \to \infty$. One can now observe that the parenthetical term in (142) vanishes as $\tilde{\delta}$ goes to 0, so choosing $\tilde{\delta} \in (0, 1/8)$ such that this parenthetical term is less than $\frac{\delta}{2}$ completes the proof. ∎

## APPENDIX III

## TECHNICAL LEMMAS

This appendix contains a series of self-contained technical lemmas that are included here for completeness. It is likely that many of them exist elsewhere in the literature, but we were unable to find the references.

### A. A Markov Property

*Lemma 13:* Let $A_j$ be i.i.d. Bernoulli-$p$ random variables such that $p \in (0, 1)$. Define stopping times

$$\tilde{N}_{2k-1} = r \cdot \inf\{i > r^{-1}\tilde{N}_{2k-2} : \sum_{j=1}^{i} A_i - i \cdot q \geq \tau\} \ , \tag{143}$$

$$\tilde{N}_{2k} = r \cdot \inf\{i > r^{-1}\tilde{N}_{2k-1} : \sum_{j=1}^{i} A_i - i \cdot q < \tau\} \ , \tag{144}$$

where $\tilde{N}_0 = 0$. Then for all real $\tau, q$ and integers $r$, and on the event $\{\tilde{N}_{2k} < \infty\}$,

$$\mathbb{P}(\tilde{N}_{2k+1} - \tilde{N}_{2k} > \ell \cdot r | \tilde{N}_{2k} < \infty) = \mathbb{P}(\tilde{N}_1 \geq \ell \cdot r | \tau - (q+1) \cdot r \leq \tilde{S}_0 < \tau) \ . \tag{145}$$

*Proof:* Define $\tilde{S}_i = \sum_{j=1}^{i} A_i - i \cdot q \cdot r$ and note that it is Markov.

$\mathbb{P}(\tilde{N}_{2k+1} - \tilde{N}_{2k} > \ell \cdot r | \tilde{N}_{2k} < \infty)$

$$= \mathbb{P}(\tilde{S}_{\tilde{N}_{2k}+1} < \tau | \tau - (q+1) \cdot r \leq \tilde{S}_{\tilde{N}_{2k}} < \tau, \tilde{N}_{2k} < \infty) \prod_{m=1}^{\ell \cdot r - 1} \mathbb{P}(\tilde{S}_{\tilde{N}_{2k}+m+1} < \tau | \tilde{S}_{\tilde{N}_{2k}+m} < \tau, \tilde{N}_{2k} < \infty) \tag{146}$$

$$= \mathbb{P}(\tilde{S}_1 < \tau | \tau - (q+1) \cdot r \leq \tilde{S}_0 < \tau, \tilde{N}_{2k} < \infty) \prod_{m=1}^{\ell \cdot r - 1} \mathbb{P}(\tilde{S}_{m+1} < \tau | \tilde{S}_m < \tau) \tag{147}$$

$$= \mathbb{P}(\tilde{N}_1 \geq \ell \cdot r | \tau - (q+1) \cdot r \leq \tilde{S}_0 < \tau) \tag{148}$$

where (146) follows from the definition of the stopping times and the fact that $\tilde{S}_i$ is Markov, (147) from the strong Markov property [21, p. 285, Theorem 5.2.4], and (148) by the definition of the stopping time.

∎

*B. Hypothesis Testing and Codebook Errors*

*Lemma 14:* Let $\mathcal{C}$ be a random codebook for the DMC $p_{Y|X}(y|x)$ with $2^{nR}$ codewords where the codewords are generated independently according to the distribution

$$p_{X^\ell}(x^n) = \prod_{i=1}^{\ell} p_X(x_i) \ . \tag{149}$$

Let $C = I(\mathbf{X}; \mathbf{Y})$. If maximum likelihood decoding is used at the decoder, then for $R < C$, the error probability $P_{\text{error}}$ is bounded from above by

$$\mathbb{P}(W \neq \hat{W}) \leq e^{-\ell \frac{(C-R)^2}{8/e^2 + 4(\log |\mathcal{Y}|)^2}} \ . \tag{150}$$

*Proof:* The result is based on an exercise in Gallager's book [24, p. 539, Problem 5.23], which in turn derives from a result in that text [24, p. 138, Theorem 5.6.2]. While there is an error in the derivation outlined in that exercise, a corrected proof is given in [25]. ∎

*Lemma 15:* Let $p_1(y)$ and $p_2(y)$ be two probability distributions such that $p_1(y) \neq p_2(y)$ for at least one $y \in \mathcal{Y}$. Given $\ell$ samples of one of these distributions, then there exists a (random) hypothesis test $T$ such that

$$P(T(Y^\ell) \neq i | Y^\ell \text{ selected iid from } p_i) \leq e^{-\ell \cdot r} \tag{151}$$

where $r > 0$ is some constant.

*Proof:* The following is simply a variation on Stein's lemma. We will construct a randomized hypothesis test that gives this performance. For each sample, independently choose with probability $\lambda \in (0, 1)$ a symbol uniformly over the alphabet; with probability $1 - \lambda$, choose the sample. This generates a new sequence of independent random variables $\tilde{Y}_i$ with distribution

$$\tilde{p}_i(y) = (1 - \lambda)p_i(y) + \frac{\lambda}{|\mathcal{Y}|} \ . \tag{152}$$

Note that our assumptions imply that $\tilde{p}_1(y) \neq \tilde{p}_2(y)$ for at least one $y \in \mathcal{Y}$, and thus

$$D(\tilde{p}_1 \| \tilde{p}_2) > 0, D(\tilde{p}_2 \| \tilde{p}_1) > 0 \ . \tag{153}$$

Now observe that

$$\log \frac{\tilde{p}_2(y)}{\tilde{p}_1(y)} \leq \log \frac{|\mathcal{Y}|}{\lambda} \ , \tag{154}$$

$$\log \frac{\tilde{p}_1(y)}{\tilde{p}_2(y)} \leq \log \frac{|\mathcal{Y}|}{\lambda} \ . \tag{155}$$

Using this, we construct the random hypothesis test $T$ as follows. Let

$$Z_i = \frac{1}{\log \frac{|\mathcal{Y}|}{\lambda}} \cdot \log \frac{\tilde{p}_1(\tilde{Y}_i)}{\hat{p}_1(\tilde{Y}_i)} . \tag{156}$$

We now define our hypothesis test to be

$$T(Y^\ell) = \begin{cases} 1 & \sum_{i=1}^{\ell} Z_i \geq 0 \\ 2 & \text{otherwise} \end{cases} . \tag{157}$$

If $\sum_{i=1}^{\ell} Z_i \geq 0$, $T$ maps to 1. Otherwise, it maps to 2. When distribution 2 is the true one, we express the error probability as

$$\mathbb{P}(\sum_{i=1}^{\ell} Z_i \geq 0 | \text{selected from } p_2) = \mathbb{P}(\ell^{-1} \sum_{i=1}^{\ell} Z_i + D(\tilde{p}_2 \| \tilde{p}_1) \geq D(\tilde{p}_2 \| \hat{p}_1) | \text{selected from } p_2) . \tag{158}$$

By Hoeffding's inequality [22, p. 57, Corollary 2.4.7], we can bound this probability by

$$\mathbb{P}(\ell^{-1} \sum_{i=1}^{\ell} Z_i + D(\tilde{p}_2 \| \tilde{p}_1) \geq D(\tilde{p}_2 \| \hat{p}_1) | \text{selected from } p_2) \leq e^{-\ell \left( \frac{D(\tilde{p}_2 \| \hat{p}_1)}{\log \frac{|\mathcal{Y}|}{\lambda}} \right)^2 / 2} . \tag{159}$$

By a similar argument, one can also show that

$$\mathbb{P}(\ell^{-1} \sum_{i=1}^{\ell} Z_i - D(\tilde{p}_1 \| \tilde{p}_2) \leq -D(\tilde{p}_1 \| \hat{p}_2) | \text{selected from } p_1) \leq e^{-\ell \left( \frac{D(\tilde{p}_1 \| \hat{p}_2)}{\log \frac{|\mathcal{Y}|}{\lambda}} \right)^2 / 2} . \tag{160}$$

Setting $r = \frac{1}{2} \left( \min \left\{ \frac{D(\tilde{p}_1 \| \hat{p}_2)}{\log \frac{|\mathcal{Y}|}{\lambda}}, \frac{D(\tilde{p}_2 \| \hat{p}_1)}{\log \frac{|\mathcal{Y}|}{\lambda}} \right\} \right)^2$ completes the proof. ∎

### C. Monotonicity, Concavity, and Continuity of a Cost-Constrained Capacity

*Lemma 16:* Let $0 \leq \epsilon_x \leq 1$ for all $x$ and define $\epsilon_0 = \min_x \epsilon_x$, which is achieved uniquely by some $x$. Then

$$C(\vec{\epsilon}, \lambda) = \max_{\substack{p(x): \\ \sum_x \epsilon_x p(x) \leq \lambda}} I(X; Y) , \tag{161}$$

is nondecreasing concave in $\lambda$ on the interval $[\epsilon_0, 1]$.

*Proof:* Since increasing $\lambda$ increases the set of channel input distributions over which to maximize, it is clear that $C(\vec{\epsilon}, \lambda)$ is nondecreasing. As a convenient shorthand, let $I_p = I(X; Y)$ denote the mutual information with the input distribution $p$. Let $p_1$ be the maximizing input distribution for $C(\vec{\epsilon}, \lambda_1)$ and $p_2$ the maximizing input distribution for $C(\vec{\epsilon}, \lambda_2)$, both of which are guaranteed to exist for $\lambda_i \in [\epsilon_0, 1]$, $i \in \{1, 2\}$. Then

$$(1 - \rho) C(\vec{\epsilon}, \lambda_1) + \rho C(\vec{\epsilon}, \lambda_2) = (1 - \rho) I_{p_1} + \rho I_{p_2} \tag{162}$$

$$\leq I_{(1-\rho)p_1 + \rho p_2} , \tag{163}$$

$$\leq C(\vec{\epsilon}, (1 - \rho)\lambda_1 - \rho\lambda_2) , \tag{164}$$

where (163) follows from the concavity of mutual information with respect to its input distribution and (164) by definition. Thus, the function is concave. ∎

*Lemma 17:* Let $0 \leq \epsilon_x \leq 1$ for all $x$ and define $\epsilon_0 = \min_x \epsilon_x$, which is achieved uniquely by some $x$. Consider

$$C(\vec{\epsilon}, \lambda) = \max_{\substack{p(x): \\ \sum_x \epsilon_x p(x) \leq \lambda}} I(X; Y) \ , \tag{165}$$

where $\lambda \in [\epsilon_0, 1]$ and $C(\vec{\epsilon}, \lambda) = 0$ for $\lambda < \epsilon_0$. Then for $0 < \Delta \leq \frac{1}{4}$,

$$0 \leq C(\vec{\epsilon}, \lambda + \Delta) - C(\vec{\epsilon}, \lambda) \leq -6\Delta \log \frac{2\Delta}{|\mathcal{X}| \cdot |\mathcal{Y}|} \ . \tag{166}$$

*Proof:* The lower bound follows immediately from Lemma 16. For the upper bound, note that $C(\vec{\epsilon}, \epsilon_0) = 0$ since the constraint can only be met by applying all the probability to a single choice of $x$. Let $\lambda \geq \epsilon_0$. Let $p_1(x)$ be the maximizing input distribution for $C(\vec{\epsilon}, \lambda + \Delta)$. If $p_1(x)$ is found in the set of valid input distributions for $C(\vec{\epsilon}, \lambda)$, then $C(\vec{\epsilon}, \lambda + \Delta) = C(\vec{\epsilon}, \lambda)$. Otherwise, we can define $p_2(x) = \sum_x \epsilon_x p_2(x) = \lambda$ and observe that

$$\lambda \leq \sum_x \epsilon_x p_1(x) \leq \lambda + \Delta \ , \tag{167}$$

which implies that

$$0 \leq \sum_x \epsilon_x (p_1(x) - p_2(x)) \leq \Delta \tag{168}$$

$$0 \leq \sum_x (1 - \epsilon_x)(p_2(x) - p_1(x)) \leq \Delta \ . \tag{169}$$

Thus,

$$\sum_x |p_2(x) - p_1(x)| = \sum_x \epsilon_x |p_2(x) - p_1(x)| + \sum_x (1 - \epsilon_x)|p_2(x) - p_1(x)| \tag{170}$$

$$\leq 2\Delta \tag{171}$$

Let $I_p = I(X; Y)$ when the input distribution for $X$ is $p$. Then by the continuity of entropy [23, Lemma 2.7, p. 33],

$$C(\vec{\epsilon}, \lambda + \Delta) = I_{p_1} \tag{172}$$

$$= I_{p_2} + (I_{p_1} - I_{p_2}) \tag{173}$$

$$\leq C(\vec{\epsilon}, \lambda) - 3 \cdot 2\Delta \log \frac{2\Delta}{|\mathcal{X}| \cdot |\mathcal{Y}|} \tag{174}$$

This is still valid as an upper bound for $\lambda < \epsilon_0$ because of the monotonicity of $C(\vec{\epsilon}, \lambda)$ from Lemma 16. ∎

*Lemma 18:* Let $0 \leq \epsilon_x \leq 1$ for all $x$ and define $\epsilon_0 = \min_x \epsilon_x$, which is achieved uniquely by some $x$, so $\epsilon_0 < \epsilon_1 = \max_x \epsilon_x$. Consider

$$C(\vec{\epsilon}, \lambda) = \max_{\substack{p(x): \\ \sum_x \epsilon_x p(x) \leq \lambda}} I(X; Y) \ , \tag{175}$$

where $\lambda \in [\epsilon_0, 1]$ and $C(\vec{\epsilon}, \lambda) = 0$ for $\lambda < \epsilon_0$. Furthermore,

$$\sum_x |\epsilon_x - \tilde{\epsilon}_x| \leq \Delta \leq \frac{1}{4} \ . \tag{176}$$

Then

$$|C(\vec{\epsilon}, \lambda) - C(\vec{\tilde{\epsilon}}, \lambda)| \leq -6\Delta \log \frac{2\Delta}{|\mathcal{X}| \cdot |\mathcal{Y}|} \ . \tag{177}$$

*Proof:* The constraint on $p(x)$ in $C(\vec{\tilde{\epsilon}}, \lambda)$ can be rewritten as

$$\sum_x \epsilon_x p(x) \leq \lambda - \sum_x (\tilde{\epsilon}_x - \epsilon_x) p(x) \ . \tag{178}$$

Since $\sum_x (\tilde{\epsilon}_x - \epsilon_x) p(x) \leq \sum_x |\tilde{\epsilon}_x - \epsilon_x|$, a tighter constraint on $p(x)$ is $\sum_x \epsilon_x p(x) \leq \lambda - \Delta$, and since $\sum_x (\tilde{\epsilon}_x - \epsilon_x) p(x) \geq -\sum_x |\tilde{\epsilon}_x - \epsilon_x|$, a looser constraint on $p(x)$ is $\sum_x \epsilon_x p(x) \leq \lambda + \Delta$. Thus

$$C(\vec{\epsilon}, \lambda - \Delta) \leq C(\vec{\tilde{\epsilon}}, \lambda) \leq C(\vec{\epsilon}, \lambda + \Delta) \ . \tag{179}$$

Then one can write

$$C(\vec{\epsilon}, \lambda) - C(\vec{\tilde{\epsilon}}, \lambda)$$

$$= C(\vec{\epsilon}, \lambda) - C(\vec{\epsilon}, \lambda - \Delta) + C(\vec{\epsilon}, \lambda - \Delta) - C(\vec{\tilde{\epsilon}}, \lambda) \tag{180}$$

$$\leq C(\vec{\epsilon}, \lambda) - C(\vec{\epsilon}, \lambda - \Delta) \ , \tag{181}$$

and similarly,

$$C(\vec{\epsilon}, \lambda) - C(\vec{\tilde{\epsilon}}, \lambda)$$

$$= C(\vec{\epsilon}, \lambda) - C(\vec{\epsilon}, \lambda + \Delta) + C(\vec{\epsilon}, \lambda + \Delta) - C(\vec{\tilde{\epsilon}}, \lambda) \tag{182}$$

$$\geq C(\vec{\epsilon}, \lambda) - C(\vec{\epsilon}, \lambda + \Delta) \ , \tag{183}$$

Lemma 17 then implies that

$$|C(\vec{\epsilon}, \lambda) - C(\vec{\tilde{\epsilon}}, \lambda)| \leq -6\Delta \log \frac{2\Delta}{|\mathcal{X}| \cdot |\mathcal{Y}|} \ . \tag{184}$$

∎

## References

[1] J. Mitola, "Cognitive radio: An integrated agent architecture for software defined radio," Ph.D. dissertation, Royal Institute of Technology, Sweden, 2000.

[2] R. Tandra and A. Sahai, "Fundamental limits on detection in low SNR under noise uncertainty," in *WirelessCom 05 Symposium on Signal Processing*, Maui, Hawaii, Jun. 2005.

[3] A. Sahai, N. Hoven, and R. Tandra, "Some fundamental limits on cognitive radio," in *Allerton Conference on Communication, Control, and Computing*, Monticello, Illinois, Oct. 2004.

[4] R. Tandra and A. Sahai, "SNR walls for signal detection," *IEEE Journal on Selected Topics in Signal Processing*, vol. 2, no. 1, pp. 4–17, February 2008.

[5] N. Devroye, P. Mitran, and V. Tarokh, "Achievable rates in cognitive radio channels," *IEEE Transactions on Information Theory*, vol. 52, no. 5, pp. 1813–1827, May 2006.

[6] A. Jovičić and P. Viswanath, "Cognitive radio: An information-theoretic perspective," 2006. [Online]. Available: http://www.citebase.org/abstract?id=oai:arXiv.org:cs/0604107

[7] I. Maric, A. Goldsmith, G. Kramer, and S. Shamai(Shitz), "On the capacity of interference channels with a cognitive transmitter," in *2007 Workshop on Information Theory and Applications (ITA)*, UCSD, La Jolla, CA, January 2007.

[8] ——, "On the capacity of interference channels with a partially-cognitive transmitter," in *2007 IEEE International Symposium on Information Theory*, Nice, France, June 2007.

[9] I. Maric, R. Yates, and G. Kramer, "Capacity of interference channels with partial transmitter cooperation," *IEEE Transactions on Information Theory*, vol. 53, no. 10, pp. 3536–3548, October 2007.

[10] A. Goldsmith, S. Jafar, I. Maric, and S. Srinivasa, "Breaking spectrum gridlock with cognitive radios: An information theoretic perspective," *Proceedings of the IEEE*, 2008.

[11] N. Devroye, M. Vu, and V. Tarokh, "Achievable rates and scaling laws in cognitive radio channels," *EURASIP Journal on Wireless Communications and Networking, special issue on Cognitive Radio and Dynamic Spectrum Sharing Systems*, 2008.

[12] R. Ahlswede, "The capacity region of a channel with two senders and two receivers," *Annals of Probability*, vol. 2, no. 5, pp. 805–814, 1974.

[13] H. Sato, "Two-user communication channels," University of Hawaii, Honolulu, HI, Tech. Rep. B75-29, October 1975.

[14] A. Carleial, "Interference channels," *IEEE Transactions on Information Theory*, vol. 24, no. 1, pp. 60–70, January 1978.

[15] M. H. M. Costa, "On the Gaussian interference channel," *IEEE Transactions on Information Theory*, vol. 31, pp. 607–615, 1985.

[16] M. Gastpar, "On capacity under receive and spatial spectrum-sharing constraints," *IEEE Transactions on Information Theory*, vol. 53, no. 2, pp. 471–487, February 2007.

[17] A. Ghasemi and E. Sousa, "Capacity of fading channels under spectrum-sharing constraints," in *IEEE International Conference on Communications*, vol. 10, June 2006, pp. 4373–4378.

[18] R. Mudumbai, B. Wild, U. Madhow, and K. Ramchandran, "Distributed beamforming using 1 bit feedback: From concept to realization," in *Allerton Conference on Communication, Control, and Computing*, Monticello, Illinois, September 2006.

[19] R. Mudumbai, J. Hespanha, U. Madhow, and G. Barriac, "Scalable feedback control for distributed beamforming in sensor networks," in *International Symposium on Information Theory*, Adelaide, Australia, September 2005.

[20] T. Cover and J. Thomas, *Elements of Information Theory*, 2nd ed.   Wiley-Interscience, 2006.

[21] R. Durrett, *Probability: Theory and Examples, 3rd ed*, ser. Information and System Sciences Series.   Duxbury, 2004.

[22] A. Dembo and O. Zeitouni, *Large Deviations Techniques and Applications, 2nd ed*, ser. Stochastic Modelling and Applied Probability.   Springer, 1998.

[23] I. Csiszár and J. G. Körner, *Information Theory: Coding Theorems for Discrete Memoryless Systems*.   Orlando, FL, USA: Academic Press, Inc., 1982.

[24] R. Gallager, *Information Theory and Reliable Communication*.   John Wiley and Sons, 1968.

[25] C. Chang, unpublished manuscript, 2006.